\documentclass[pdflatex,sn-mathphys-num]{sn-jnl}

\usepackage{graphicx}%
\usepackage{multirow}%
\usepackage{amsmath,amssymb,amsfonts}%
\usepackage{amsthm}%
\usepackage{mathrsfs}%
\usepackage[title]{appendix}%
\usepackage{xcolor}%
\usepackage{textcomp}%
\usepackage{manyfoot}%
\usepackage{booktabs}%
\usepackage{algorithm}%
\usepackage{algorithmicx}%
\usepackage{algpseudocode}%
\usepackage{listings}%

\usepackage{physics}
\usepackage{caption} 

\theoremstyle{thmstyleone}%
\theoremstyle{thmstyletwo}%

\theoremstyle{thmstylethree}%

\usepackage{bm}

\begin{document}


\title{A multiple occupancy cell fluid model with competing attraction and repulsion interactions}


\author*{\fnm{R.~V.} \sur{Romanik}}\email{romanik@icmp.lviv.ua}

\author{\fnm{O.~A.} \sur{Dobush}}\email{dobush@icmp.lviv.ua}

\author{\fnm{M.~P.} \sur{Kozlovskii}}\email{mpk@icmp.lviv.ua}

\author{\fnm{I.~V.} \sur{Pylyuk}}\email{piv@icmp.lviv.ua}

\author{\fnm{M.~A.} \sur{Shpot}}\email{shpot.mykola@gmail.com}
\affil{\orgname{Yukhnovskii Institute for Condensed Matter Physics of the National Academy of Sciences of Ukraine}, \orgaddress{
		\city{Lviv}, \postcode{79011}, \country{Ukraine}}}


\abstract{An analytically solvable cell fluid model with unrestricted cell occupancy, infinite‑range Curie–Weiss–type attraction and short‑range intra‑cell repulsion is studied within the grand‑canonical ensemble. Building on an exact single‑integral representation of the grand partition function, we apply Laplace’s method to obtain asymptotically exact expressions for the pressure, density and equation of state. The model exhibits a hierarchy of first‑order transitions, each terminating at a critical point. We determine the coordinates of the first five such points. Recasting the formalism in dimensionless variables highlights the explicit temperature dependence of all thermodynamic functions. This enables us to derive a closed‑form expression for the entropy. The results reveal pronounced entropy minima around integer cell occupancies and reproduce density‑anomaly isotherm crossings analogous to those in core‑softened models.}




\maketitle

\section{Introduction}\label{sec1}
Lattice models have played an important role in statistical physics for a long time,
offering simplified yet physically meaningful representations of many-particle
systems. Despite the spatial discretisation they impose,
lattice models can qualitatively reproduce the essential thermodynamic
properties of fluids and offer convenience for numerical simulations~\cite{Panagiotopoulos00}. The most extensively studied lattice gas model is the one isomorphic to the Ising model, in which each lattice site may be either occupied by a single particle or left empty~\cite{LY52}. In contrast, models that allow more than one particle per site have been studied less extensively, even though they can show equally or even more interesting physical behavior. A notable example is the double-occupancy lattice gas, isomorphic to the Blume-Capel model, recently analyzed in~\cite{LYZ21,WLWLX24}. More generally, models allowing arbitrary multiple occupancy have been explored in~\cite{LWL98,FHL04,FL18}.
The models of this kind possess richer phase behavior compared to the single-occupancy lattice gas. Similar features arise in quantum many-body physics, where lattice-boson models
are inherently multiple-occupancy systems~\cite{FWGF89,SJS12,KS19}.

As emphasized by Stanley~\cite[Appendix~A]{Stanley71}, it is often more
physically transparent to speak of the occupancy of cells by the centers of particles, rather than restrict the atoms' positions to a discrete set of sites in a lattice. This perspective underpins the usage of the `cell' terminology, rather than the `lattice' one~\cite{Barker54,DMMR05,Rebenko13}.
Within this framework, one can vary both the maximal occupancy and the range of inter-particle interactions. Extending the interaction to infinite range drives the system to mean-field behavior, and often allows for an exact solution~\cite{Baxter82,BCC06}

A cell model that combines unrestricted cell occupancy with an infinite-range Curie-Weiss-type attraction was introduced in~\cite{KK16,KKD20} and \cite{KKD18}.
The main features of the model are as follows:
(i) each cell may contain an arbitrary number of particles;
(ii) a global attractive interaction is imposed between all particle pairs in the system, regardless of their positions;
(iii) a local repulsive interaction acts pairwise between particles within the same cell;
(iv) the thermodynamic stability of the system is ensured by assuming that the repulsion dominates over the attraction, preventing particle collapse or unlimited clustering.
In the thermodynamic limit, the model's grand partition function can be calculated asymptotically exactly~\cite{KKD20}, revealing a hierarchy of critical points analogous to those of the generalized Curie-Weiss magnet~\cite{EE88}.

The goal of this study is twofold. First, we revisit the known results for the cell model with Curie-Weiss interaction and rewrite them in the notation standard to the theory of simple liquids~\cite{HansenMcDonald13}, thereby making the temperature dependence of
thermodynamic quantities explicit. Second, we derive a closed-form expression for the entropy, a quantity left unexplored in previous treatments, and analyze its behavior near the multiple critical points.

The paper is organized as follows. In Section~\ref{sec:model}, we formulate the model, establish the notation to be used throughout the paper, and present the expression for the grand partition function. In Section~\ref{sec:thermmodynamics}, the explicit expressions are derived for the pressure and the average number of particles as functions of temperature and chemical potential, as well as for the equation of state.
Section~\ref{sec:CP} focuses on the details of calculating the critical points' coordinates.
In Section~\ref{sec:entropy}, we proceed with the calculation of the entropy.
Finally, summary and conclusions are present in Section~\ref{sec:conclusion}.

\section{Model}\label{sec:model}
We begin by introducing the model and notation, and then summarize the main known results.
In this section, we closely follow the argument of \cite{KKD20}. Subsequently, we extend its framework by introducing dimensionless physical quantities as standard in the theory of fluids~\cite{HansenMcDonald13}.

\bigskip
\textbf{General setting.}
The statistical mechanics model defined below is studied within the formalism of the grand canonical ensemble (see e.g. \cite[Sec. 2.4]{HansenMcDonald13}). An open system of point particles is considered in volume $V\subset\mathbb R^3$ in three space dimensions. The total volume $V$ is partitioned into $N_v$ non-overlapping congruent cubic cells $\Delta_l$, $l\in\{1,...,N_v\}$, each of volume $v$, such that $V$ is the union of all $\Delta_l$'s:
\begin{equation}\label{volume}
	V = \bigcup_{l=1}^{N_v}\Delta_l,\qquad
\Delta_l \cap \Delta_{l'} = \emptyset, \text{ if } l \neq l',\qquad\mbox{and}\quad
V = v N_v.
\end{equation}

\bigskip
\textbf{Interaction.} Consider a configuration $\gamma_N= \{\vb{r}_1, ..., \vb{r}_N\}$ of particles contained in the volume $V$,
where $\vb{r}_i$ is the position vector of the $i$-th particle and $\abs{\gamma_N}\equiv N$ is the number of particles in this configuration.
In fact, for any given number $N$ of particles, there are infinitely many associated configurations $\gamma_N$ each corresponding to a specific random distribution of these particles over $V$ and, thus, over cells $\Delta_l$.
The total potential energy of interaction between particles in a given configuration $\gamma_N$ is defined as
\begin{eqnarray}\label{WN}
	W_{N_v}(\gamma_N)=\frac{1}{2} \sum_{\vb{r}_i,\vb{r}_j \in \gamma_N} \Phi_{N_v} (\vb{r}_i,\vb{r}_j)=\frac12 {\sum_{i=1}^N \sum_{j=1}^N} \Phi_{N_v} (\vb{r}_i,\vb{r}_j),
\end{eqnarray}
where $\Phi_{N_v}=\Phi_{N_v}(\vb{r}_i, \vb{r}_j)$ is given by the two-point interaction in the form
\begin{equation}
	\label{def:curie-weiss-pot}
	\Phi_{N_v}(\vb{r}_i, \vb{r}_j) = -\frac{J_1}{N_v} + J_2\sum_{l=1}^{N_v} \mathbb{I}_{\Delta_l}(\vb{r}_i) \mathbb{I}_{\Delta_l}(\vb{r}_j).
\end{equation}
The first term in $\Phi_{N_v}$ describes a global Curie-Weiss (mean-field-like) attraction for any pair of particles in the system.
The strength of this attraction is controlled by an energy parameter $J_1 > 0$. The second term in $\Phi_{N_v}$ describes a local repulsion between two particles contained within the same cell $\Delta_l$ and is characterized by the parameter $J_2 > 0.$
Here, $\mathbb{I}_{\Delta_l}(\vb{r})$ is the indicator function of the cell $\Delta_l$,
\begin{equation}
	\label{def:I}
	\mathbb{I}_{\Delta_l} (\vb{r}) = \left\{
	\begin{array}{ll}
		1, \quad \vb{r} \in \Delta_l,
		\\
		0, \quad \vb{r} \notin \Delta_l.
	\end{array}
	\right.
\end{equation}
Therefore, two cases are possible for $\Phi_{N_v}(\vb{r}_i, \vb{r}_j)$, depending on whether two particles belong to the same cell or to different ones. Explicitly, one has
\begin{equation}\label{POS}
	\Phi_{N_v}(\vb{r}_i, \vb{r}_j) = \left\{
	\begin{array}{ll}
		-J_1/N_v+ J_2, & \text{when particles are in the same cell,}
		\\
		-J_1/N_v, & \text{otherwise.}
	\end{array}
	\right.
\end{equation}

The sum over all particle pairs in the total energy $W_{N_v}(\gamma_N)$ brings about a competition between the global attractive energy of particles and their repulsive energy at ``short'' distances. This competition leads to a certain kind of frustration in the system, which is responsible for the non-trivial phase behavior discussed in \cite{KD22}.

Note that the self-interaction terms $\Phi_{N_v}(\vb{r}_i,\vb{r}_i)$ are included in $W_{N_v}$ from \eqref{WN}. This does not affect the physics of the model. Given that $\sum_{\vb{r}_i\in\gamma_N}1=\abs{\gamma_N}=N$, we combine the definitions \eqref{def:curie-weiss-pot} and \eqref{WN}, which yields
\begin{equation}\label{WNS}
W_{N_v}(\gamma_N)=-\frac{J_1}{2N_{v}} N^2+\frac{J_2}{2} \sum_{\vb{r}_i,\vb{r}_j \in \gamma_N} \sum_{l=1}^{N_v} \mathbb{I}_{\Delta_l}(\vb{r}_i)\mathbb{I}_{\Delta_l}(\vb{r}_j)\,.
\end{equation}

Let us notice the different character of summations over positions $\vb{r}_i$ and $\vb{r}_j$
in two terms present in \eqref{WNS}.
The first term in~\eqref{WNS} is the result of the double summation $\sum_{\vb{r}_i,\vb{r}_j \in \gamma_N}$ applied to a constant, which evaluates to $N^2$ and does not depend on any specific distribution of points $\vb{r}_i$ within the configuration $\gamma_N$.
The situation is different with the second term. Here, the result of summation over $\bm r_i$ and $\vb{r}_j$ heavily depends on the distribution of particles with these coordinates over the cells constituting the volume $V$.
According to the rule \eqref{POS}, the number of appearances of $J_2$ directly depends on the number of particles and, consequently, on their pairs, occupying each cell $\Delta_l$ in the volume $V$.

\bigskip
\textbf{Stability.} For thermodynamic stability of the system with two-point interactions
$\Phi_{N_v}(\vb{r},\vb{r}')$, the rigorous inequality~\cite{KKD20,Ruelle70}
\begin{equation}\label{RUE}
\int_{V} \Phi_{N_v}(\vb{r},\vb{r}'){\rm d} \vb{r}' > 0
\end{equation}
must hold for any $\vb{r}\in V$.
Taking into account \eqref{def:curie-weiss-pot} and \eqref{POS}, we
see that there is one contribution to this integral from a cell containing the coordinate $\vb{r}$, and $(N_v - 1)$ contributions from other cells:
\begin{eqnarray*}
	\int_{V} \Phi_{N_v}(\vb{r},\vb{r}') {\rm d} \vb{r}' & = & \left(-\frac{J_1}{N_v} + J_2\right)v - (N_v - 1)\,\frac{J_1}{N_v}\,v=(J_2 - J_1)v.
\end{eqnarray*}
Thus, the requirement \eqref{RUE} implies that the energies of repulsion and attraction interactions must obey the inequality
\begin{equation*}
	J_2 > J_1.
\end{equation*}
It is useful to introduce their ratio via
\begin{equation}\label{a}
a\equiv J_2/J_1>1.
\end{equation}
The quantity $a$ can be considered as one of two adjustable parameters of the studied model, another one being $v^*$ introduced below~\eqref{def:v_star}.

\bigskip
\textbf{The grand partition function} (GPF) is given by the following sum over all possible numbers $N$ of particles in volume $V$ (see e.g. \cite[(2.4.6)]{HansenMcDonald13}):
\begin{equation}\label{ZGR}
	\Xi=\sum_{N=0}^{\infty}\frac{\zeta^N}{N!}Z_N.
\end{equation}
Here $\zeta$ is the activity
\begin{equation}\label{Lambda}
	\zeta = \frac{\exp(\beta \mu)}{\Lambda^3},
\end{equation}
$\beta = 1/(k_{\rm B} T)$ is the inverse temperature, $k_{\rm B}$ is the Boltzmann constant, $T$ is the (absolute) temperature, and $\mu$ is the chemical potential.
The value $\Lambda = (2\pi\beta\hbar^2/m)^{1/2}$ is the de Broglie thermal wavelength,
$\hbar$ is the Planck constant, and $m$ is the mass of a particle.
The presence of $\Lambda$ in~\eqref{ZGR} arises from (a) the choice of the normalization factor in the definition of the partition function, ensuring consistency between classical and quantum statistical mechanics, and (b) the integration over momenta (see, e.g.,~\cite[Section~2.3]{HansenMcDonald13}).

The configuration integral $Z_N$ is defined as
\begin{equation}
	Z_N = \int\exp(-\beta W_{N_v}(\gamma_N))({\rm d}\vb{r})^N,
\end{equation}
with $({\rm d} \vb{r})^N \equiv {\rm d}{\vb r_1} \dotsc {\rm d}{\vb r_N}$.
In $Z_N$, the integration goes over all configurations $\gamma_N$ of $N$ particles inside the volume $V$.

The mean values of physical quantities are defined as statistical averages over the grand canonical ensemble via
\begin{equation}
	\langle \ldots \rangle = \frac{1}{\Xi} \sum_{N=0}^{\infty}(\ldots)\frac{\zeta^N}{N!}Z_N.
\end{equation}
Within the framework of the grand canonical ensemble, the observable average number of particles in the system is given by
\begin{equation}\label{Ndef}
\langle N \rangle=\beta^{-1}\frac{\partial\ln\Xi}{\partial\mu}\, = \frac{1}{\Xi} \sum_{N=0}^{\infty}N\frac{\zeta^N}{N!}Z_N \,.
\end{equation}
This is an important quantity for further consideration.

Taking into account the expression \eqref{WNS} for the total energy $W_{N_v}(\gamma_N)$
we write the GPF of our model in the form
\begin{equation}
	\label{eq:gpf1}
	\Xi = \sum_{N=0}^{\infty} \frac{\Lambda^{-3N}}{N!}
	\int
	\exp\left[\beta\mu N + \frac{\beta J_1}{2N_{v}} N^2 - \frac{\beta J_2}{2} \sum_{\vb{r}_i,\vb{r}_j \in \gamma} \sum_{l=1}^{N_v} \mathbb{I}_{\Delta_l}(\vb{r}_i)\mathbb{I}_{\Delta_l}(\vb{r}_j)\right]({\rm d}\vb{r})^N.
\end{equation}

In addition to introducing the cell fluid model with competing Curie-Weiss attraction and intra-cell repulsion interactions,
one of the main achievements of \cite{KKD20} was the representation of the GPF appearing in \eqref{eq:gpf1} in the form of a single integral. The corresponding result can be found in
\cite[(2.13)--(2.15)]{KKD20}. We slightly modify its derivation by taking into account the presence of the de Broglie thermal wavelength $\Lambda$ in the definition of the GPF (see \eqref{ZGR} and \eqref{Lambda}). Moreover, we express the results in terms of the
reduced dimensionless thermodynamic quantities widely accepted in the theory of simple liquids (see e.g.~\cite{HansenMcDonald13}).

We start the following section by introducing these reduced quantities. Next, we provide the explicit expressions determining the integral representation for the GPF \eqref{eq:gpf1} of the cell model.

\section{Thermodynamics}\label{sec:thermmodynamics}

Natural units for energy and length in the model under consideration are $J_1$, the Curie-Weiss attraction's strength, and $c=v^{1/3}$, the length of the cell's edge, respectively. It is a standard practice to express thermodynamic results in terms of dimensionless quantities normalized by natural units of this kind. The advantage of using such quantities is that their numerical values are typically of order of unity, which simplifies analysis. This also opens a direct possibility for comparisons with results available in the theory of liquids.

\subsection{Reduced dimensionless quantities}

For any dimensional quantity $Q$, it is convenient to introduce its dimensionless, or ``reduced'', counterpart $Q^*$. Thus, we define the following reduced quantities:
\begin{eqnarray}
	\label{def:reduced}
	T^* := \frac{k_{\rm B} T}{J_1} & \quad & \text{ -- the reduced temperature;}
	\nonumber\\
	\beta^* := \beta J_1 = \frac{1}{T^*} & \quad & \text{ -- the reduced inverse temperature;}
	\nonumber\\
	\mu^* := \frac{\mu}{J_1} & \quad & \text{ -- the reduced chemical potential;}
	\nonumber\\
	P^* := \frac{Pv}{J_1} & & \text{ -- the reduced pressure;}
	\nonumber\\
	\rho^* := \frac{\langle N \rangle}{V} v && \text{ -- the reduced density;}
	\nonumber\\
	S^* := \frac{S}{k_{\rm B} \langle N \rangle} & & \text{ -- the reduced entropy.}
\end{eqnarray}
\label{pRED}
Taking into account~\eqref{volume} and~\eqref{Ndef}, we see that the reduced density $\rho^*$ actually represents the average particle number per cell: $\rho^* = \langle N \rangle/N_v$.

Working within the grand canonical ensemble, we naturally express all physical quantities in terms of the reduced temperature $T^*$ and chemical potential $\mu^*$.

\subsection{Integral representation for the GPF}

In terms of the reduced dimensionless quantities introduced just above, the integral representation for the GPF from \eqref{eq:gpf1} is given by
(cf. \cite[(2.13)--(2.15)]{KKD20})
\begin{equation}
	\label{eq:XiInty}
	\Xi (T^*,\mu^*) = \sqrt{\frac{N_v T^*}{2\pi}} \int\limits_{-\infty}^{\infty}
\exp\left[N_v E(T^*,\mu^*; y)\right] {\rm d} y,
\end{equation}
where
\begin{equation}
	\label{def:E}
	E(T^*,\mu^*; y) = -\frac12T^*y^2 + \ln K_0(T^*,\mu^*; y),
\end{equation}
and
\begin{equation}
	\label{def:K}
	K_0(T^*,\mu^*; y) = \sum_{n=0}^{\infty} \frac{\left(v^* T^{*3/2}\right)^n}{n!} \exp[\left(y+\frac{\mu^*}{T^*}\right)n - \frac{a}{2T^*}n^2]
\end{equation}
is a special function, often emerging when studying cell models~\cite{SS05,KDR15}
The transformation of the GPF from \eqref{eq:gpf1} to \eqref{eq:XiInty} may be viewed as exact since it proceeds without any additional approximations.

A few words of explanation are in order here.
To be consistent in working with dimensionless quantities defined just above, we have expressed the de Broglie thermal wavelength $\Lambda$ (first appearing in \eqref{Lambda}) in terms of the reduced temperature via
\begin{equation}
\Lambda=\left(\frac{2\pi\beta\hbar^2}{m}\right)^{1/2}=
\left(\frac{2\pi\beta J_1\hbar^2}{m J_1}\right)^{1/2}=
\left(T^*\right)^{-1/2}\,\Lambda_J.
\end{equation}
Here, we have introduced the quantity $\Lambda_J = (2\pi\hbar^2/mJ_1)^{1/2}$ with dimension of length, and used it in defining the \emph{dimensionless} cell volume as
\begin{equation}
	\label{def:v_star}
	v^* = \frac{v}{\Lambda^3_J}\,.
\end{equation}
This entry adds to the list of dimensionless quantities \eqref{def:reduced} in the beginning of the current section.

In what follows, numerical and graphical results are presented for the following values of the model parameters:
\begin{equation}
	v^* = 5.0; \quad a = 1.2,
\end{equation}
if not specified otherwise.

Note that the presence of the de Broglie thermal wavelength $\Lambda$ in the definition \eqref{Lambda} of activity is responsible for an additional temperature dependence in the product $v^* T^{*3/2}$ in \eqref{def:K}, compared to the analogous expression in \cite[(2.15)]{KKD20}.

The integral over $y$ in \eqref{eq:XiInty} is appropriate for treatment via the Laplace method in the limit $N_v\to\infty$, which corresponds to the thermodynamic limit for the considered system.

\label{pDEF}
Quite recently, the special function $K_0$ was studied from a different, mathematical perspective in~\cite{DS24arxiv,GS25}. Moreover, it appears that it can be transformed, through a simple reparametrization, to the \textit{deformed exponential function}~\cite{SS05,Sokal12,GS25} as exemplified in \cite{GS25}.

\subsection{Laplace's method}\label{SLM}

In order to compute any standard thermodynamic quantity like pressure, particle density, entropy, we need the GPF as the starting point. In order to evaluate the GPF in the thermodynamic limit, we need to calculate the large-$N_v$ limit of the integral~\eqref{eq:XiInty} via the Laplace's method~\cite{Fedoryuk89,BenderOrszag99}.
First of all, this requires determination of the global maxima of $E(T^*,\mu^*;y)$ as the function of $y \in \mathbb{R}$.

Let $\bar{y}$ denote the extremum position of $E(T^*,\mu^*;y)$ at given temperature and chemical potential. As usual, it should obey the equation
\begin{equation}
	\label{cond:extr}
E_1(T^*,\mu^*;\bar{y}) = 0,\qquad\mbox{where}\quad
E_1(T^*,\mu^*;y)\equiv \frac{\partial}{\partial y} E(T^*,\mu^*;y).
\end{equation}

\begin{figure}[htbp]
	\includegraphics[width=0.5\textwidth,angle=0]{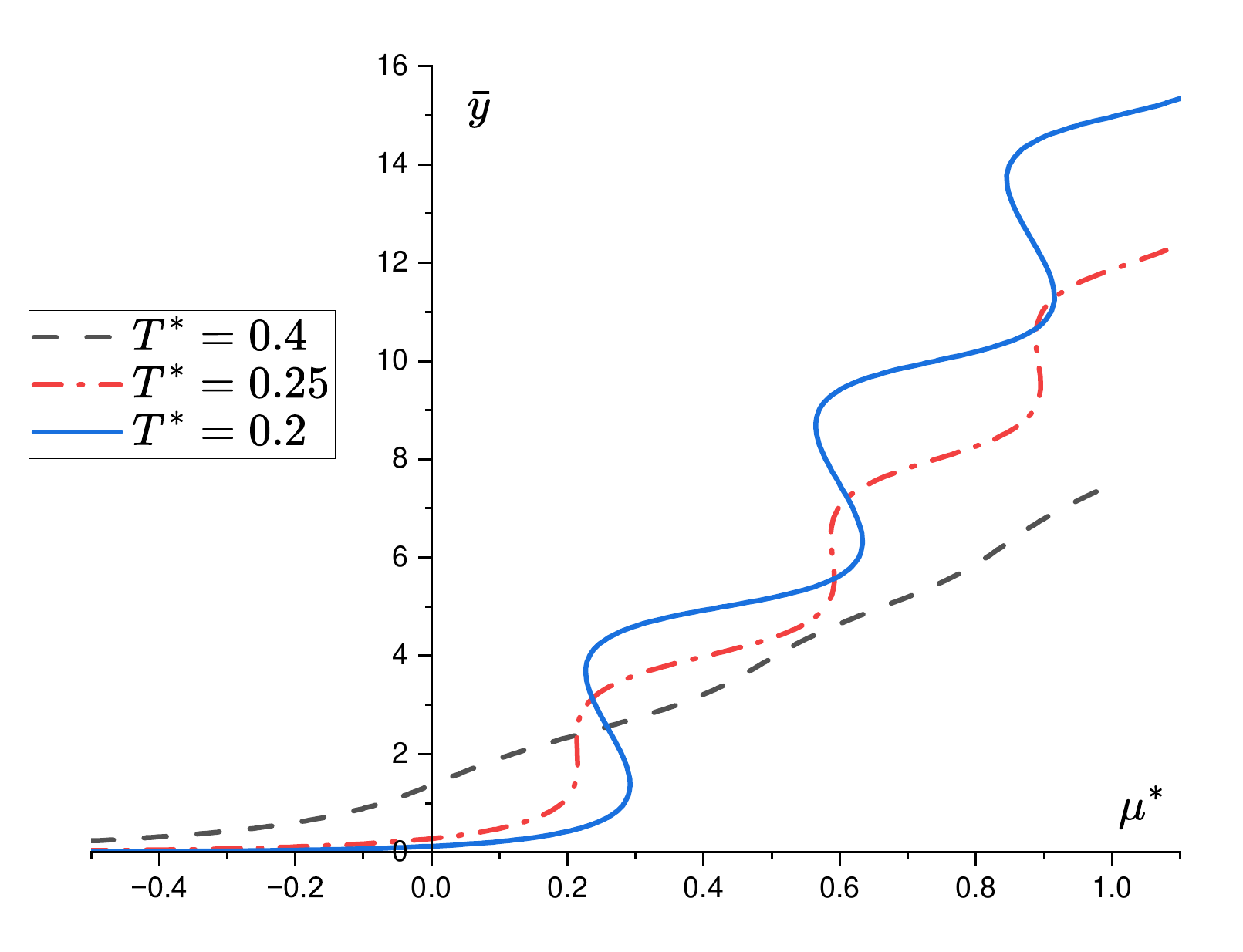}
	\hfill
	\includegraphics[width=0.5\textwidth,angle=0]{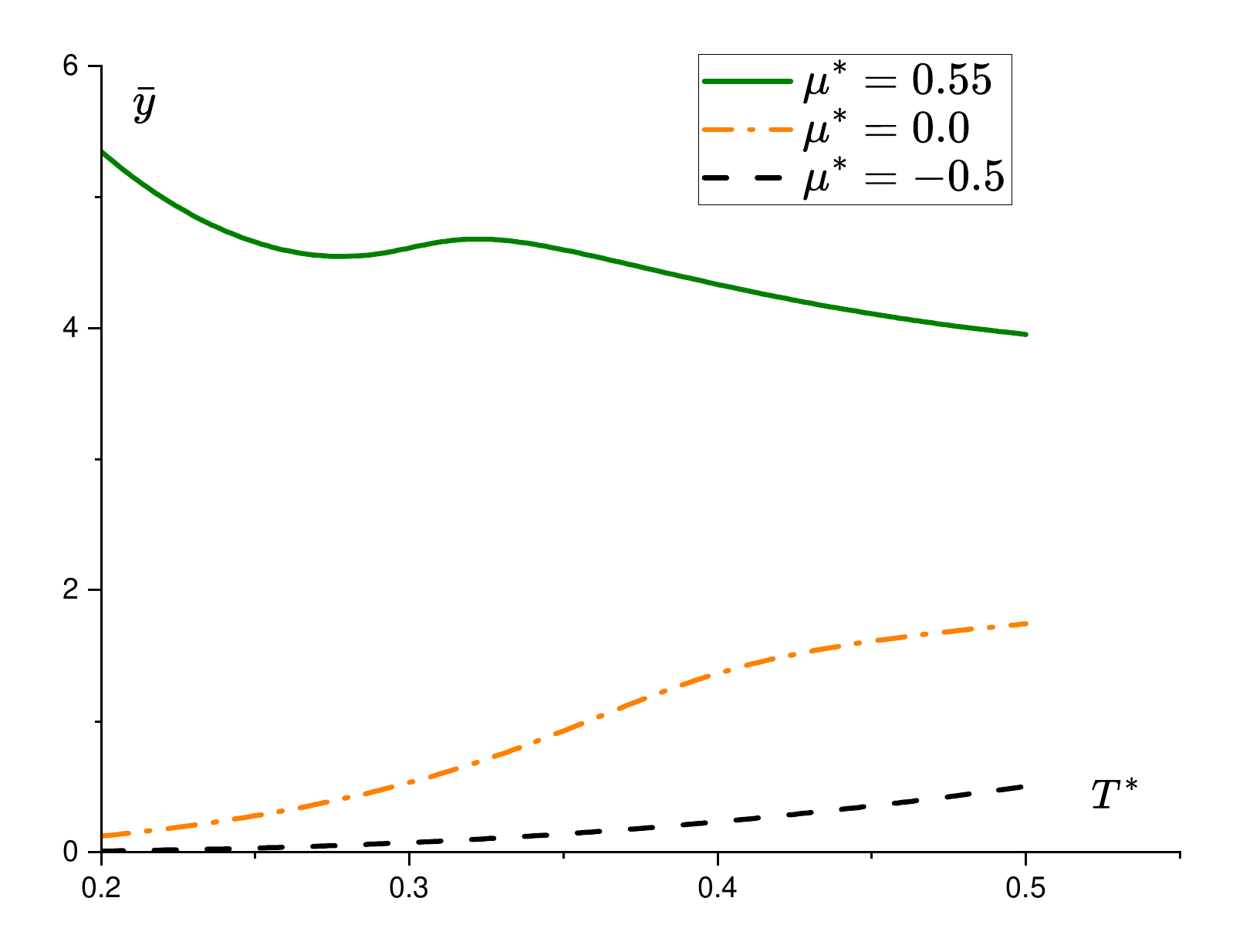}
	\\
	\parbox{0.45\textwidth}{\caption{\label{fig:YvsMu_a1} Solution $\bar{y}$ to the equation $E_1(T^*,\mu^*;\bar{y})=0$ as a function of $\mu^*$ for selected values of $T^*$. Dashed curve (black): $T^* = 0.4$; Dash-dotted curve (red): $T^* = 0.25$; Solid curve (blue): $T^*=0.2$.}}
	\hfill
	\parbox{0.45\textwidth}{\caption{\label{fig:YvsT_a2} Solution $\bar{y}$ to the equation $E_1(T^*,\mu^*;\bar{y})=0$ as a function of $T^*$ for selected values of $\mu^*$. Dashed curve (black): $\mu^* = -0.5$, Dash-dotted curve (orange): $\mu^* = 0.0$, Solid line (green): $\mu^* = 0.55$.}}
\end{figure}

By~\eqref{def:E} and~\eqref{def:K}, the explicit expression for $E_1$ is given by
\begin{align}\label{dE1}
E_1(T^*,\mu^*;y)&=-T^* y + \frac{\partial\ln K_0(T^*,\mu^*;y)}{\partial y}
\\\label{def:reducedE1}
&=-T^* y + \frac{K_1(T^*,\mu^*;y)}{K_0(T^*,\mu^*;y)},
\end{align}
and the equation for $\bar{y}$ 
reads~(cf.~\cite[(2.19)]{KKD20})
\begin{equation}
	\label{eq:bar_y}
	-T^* \bar{y} + \frac{K_1(T^*,\mu^*;\bar{y})}{K_0(T^*,\mu^*;\bar{y})} = 0.
\end{equation}
This is an implicit equation defining $\bar{y} = \bar{y}(T^*,\mu^*)$ as a function of $T^*$ and $\mu^*$.
The function $K_1$ is the first one in the hierarchy of functions defined by
\begin{eqnarray}\label{Kjd}
	K_j(T^*,\mu^*;y) & := & \frac{\partial^j}{\partial y^j} K_0(T^*,\mu^*;y)
\\\label{Kj}
	& = & \sum_{n=0}^{\infty} \frac{n^j (v^* T^{*3/2})^n}{n!} \exp[\left(y+\frac{\mu^*}{T^*}\right)n - \frac{a}{2T^*}n^2]
\end{eqnarray}
for $j=0,1,2,\ldots$ . 

As $K_0$, $K_1$, and $T^*$ are all strictly positive, the solution $\bar{y}$ to ~\eqref{eq:bar_y} is also strictly positive: $\bar{y} \in \mathbb{R}^{+}$.
Figure~\ref{fig:YvsMu_a1} illustrates the dependence of $\bar{y}$ on $\mu^*$ for different values of temperature $T^*$. When the temperature is relatively high, as in the case $T^*=0.4$, $\bar{y}$ is a single-valued function of $\mu^*$. At a low temperature, $T^*=0.2$, there are intervals in $\mu^*$ where $\bar{y}$ becomes a multi-valued function. The intermediate temperature range where $\bar{y}$ changes from being a single-valued to a multi-valued function of $\mu^*$, is of particular interest and will be examined more closely in Section~\ref{sec:CP}. Figure~\ref{fig:YvsT_a2} illustrates the dependence of $\bar{y}$ on $T^*$ for three different values of $\mu^*$. In the temperature range shown in the figure, $\bar{y}$ is an increasing function of $T^*$ for small values of $\mu^*$, but exhibits different behavior when $\mu^*$ increases to $\mu^*=0.55$. Therefore, it is also of interest to study how this crossover from one type of behavior to another occurs in the region of intermediate values of $\mu^*$.

Until this point, we did not distinguish whether the extremum point $\bar y$ corresponds to a maximum or a minimum. However, in applying the Laplace's method we have to deal with maxima of the function $E(T^*,\mu^*;y)$, for which the second derivative of this function must be negative. Thus, we have to consider the values
$\bar{y}_{\rm max}=\bar{y}_{\rm max}(T^*,\mu^*)$, for which the following two conditions
\begin{eqnarray}
	\label{cond:max}
	E_1(T^*,\mu^*; \bar{y}_{\rm max}) & = & 0,
	\nonumber\\
	E_2(T^*,\mu^*;\bar{y}_{\rm max}) & < & 0,
\end{eqnarray}
are simultaneously satisfied. Similarly as in \eqref{cond:extr}, in the second line we used a short-hand notation $E_2(T^*,\mu^*;y)$ for the second derivative of $E(T^*,\mu^*;y)$,
which is a representative of the family of derivatives
\begin{eqnarray}\label{Ej}
E_j(T^*,\mu^*;y)\equiv\frac{\partial^j}{\partial y^j} E(T^*,\mu^*;y)\qquad\mbox{with}\quad
j=0,1,2,\ldots\,.
\end{eqnarray}
In terms of functions $K_j$ from \eqref{Kj}, the function $E_2(T^*,\mu^*;y)$ reads
\begin{align}\label{dE2}
E_2(T^*,\mu^*;y)&=-T^*+\frac{\partial^2\ln K_0(T^*,\mu^*;y)}{\partial y^2}
\\\label{def:reducedE2}
&=-T^*+\frac{K_2(T^*,\mu^*;y)}{K_0(T^*,\mu^*;y)}
-\left[\frac{K_1(T^*,\mu^*;y)}{K_0(T^*,\mu^*;y)}\right]^2\!.
\end{align}
Note that for $j=3,4,\ldots$\,, the functions $E_j(T^*,\mu^*;y)$ are expressed only in terms of $K_j(T^*,\mu^*;y)$ via
\begin{equation}\label{REJ}
E_j(T^*,\mu^*;y)=\frac{\partial^j}{\partial y^j}\ln K_0(T^*,\mu^*;y)\,.
\end{equation}

It is said that $(T^*, \mu^*)$ belongs to a \textit{single-phase domain} of the half-plane
$\{(T^*,\mu^*): T^*\in\mathbb{R}^+, \mu^*\in\mathbb{R}\}$ if $E(T^*,\mu^*;y)$ has a unique global maximum $\bar{y}_{\rm max} \in \mathbb{R}^{+}$ such that conditions~\eqref{cond:max} are satisfied.

Provided that the double condition \eqref{cond:max} is satisfied, the application of the Laplace's method~(\cite[(1.21)]{Fedoryuk89}, \cite[(6.4.19c)]{BenderOrszag99}) to the integral in~\eqref{eq:XiInty} yields the result
\begin{equation}\label{laplace1}
	\Xi(T^*,\mu^*) = \left[-\frac{T^*}{E_2(T^*,\mu^*;\bar{y}_{\rm max})}\right]^{1/2}
\exp[N_v E(T^*,\mu^*;\bar{y}_{\rm max})].
\end{equation}

Typical behavior of functions $E_1(T^*,\mu^*;y)$ and $E_2(T^*,\mu^*;y)$ corresponding to the single-phase domain is shown in left panels of Figs.~\ref{fig:E1y_vs_y_a} and~\ref{fig:E2y_vs_y_a}, respectively. Further plots of these functions can be found
in other figures in Appendices~\ref{sec:figE1} and~\ref{sec:figE2}.

\subsection{Pressure}\label{sec:pres}
The pressure $P$ is given by the well-known thermodynamic relation
\begin{equation}
	\label{def:eos}
	P V = k_{\rm B} T \ln \Xi\,.  
\end{equation}
In the thermodynamic limit $N_v\to\infty$, equation \eqref{laplace1} shows that the dominant contribution to the logarithm of the GPF is given by
\begin{equation}\label{LLL}
	\lim_{N_v\to\infty}N_v^{-1}\ln\Xi(T^*,\mu^*)=E(T^*,\mu^*;\bar{y}_{\rm max})\,,
\end{equation}
which leads directly to
\begin{equation}
	\frac{P V}{k_{\rm B}T N_v} = E(T^*,\mu^*;\bar{y}_{\rm max})\,.
\end{equation}
The reduced dimensionless pressure $P^*=P^*(T^*,\mu^*)$ can therefore be expressed as
\begin{equation}
	\label{eos:reduced}
	P^* = T^* E(T^*,\mu^*;\bar{y}_{\rm max})
\end{equation}
in terms of the thermodynamic variables defined in \eqref{def:reduced}.
Note that the maximum position $\bar{y}_{\rm max}$ in \eqref{eos:reduced} is itself a function of $T^*$ and $\mu^*$: $\bar{y}_{\rm max} = \bar{y}_{\rm max}(T^*,\mu^*)$.
Since an analytical expression for $\bar{y}_{\rm max}(T^*,\mu^*)$ cannot be derived, we determine this function numerically from equation \eqref{eq:bar_y} and present the results graphically in Figures~\ref{fig:YvsMu_a1}, \ref{fig:YvsT_a2} as well as in  the following sections.

The relation \eqref{eos:reduced} actually represents the equation of state for the system under consideration, expressed in terms of thermodynamic variables $P^*$, $T^*$, and $\mu^*$. It encompasses the fact that the relevant thermodynamic behavior in the thermodynamic limit is determined by the stable global maximum of the function $E(T^*,\mu^*;y)$ from the integral representation \eqref{eq:XiInty}.

\subsection{Average number of particles}
The average particle number $\langle N \rangle$ is given by the fundamental thermodynamic relation
\begin{equation}
	\langle N \rangle = -\left(\frac{\partial \Omega}{\partial \mu}\right)_{T,V},
\end{equation}
where $\Omega = -k_{\rm B} T \ln \Xi$ is the grand potential.
Hence, using the definition \eqref{Ndef} for $\langle N \rangle$ along with the formula \eqref{def:eos} for pressure, we obtain
\begin{equation}\label{NPalt}
\frac{\langle N \rangle}{V}=\frac{\partial P}{\partial \mu}\,,
\end{equation}
where the derivative with respect to the chemical potential $\mu$ is taken at constant temperature $T$.

In terms of the reduced quantities defined in \eqref{def:reduced}, equation \eqref{NPalt} becomes
\begin{equation}\label{eq:dens}
\rho^*= \left( \frac{\partial P^*}{\partial \mu^*}\, \right)_{T^*}.
\end{equation}

To derive the reduced density $\rho^*=\rho^*(T^*,\mu^*)$ for our cell model, we differentiate \eqref{eos:reduced} using the chain rule:
\begin{equation}\label{CHR}
\rho^*=T^*\left[\frac{\partial E(T^*,\mu^*;\bar{y}_{\rm max})}{\partial\mu^*}+
\frac{\partial E(T^*,\mu^*;\bar{y}_{\rm max})}{\partial\bar{y}_{\rm max}}\,
\frac{\partial\bar{y}_{\rm max}}{\partial\mu^*}\right].
\end{equation}
The second term in \eqref{CHR} vanishes due to the extremum condition \eqref{cond:max}, which determines $\bar{y}_{\rm max}$ itself. Consequently, using the definitions \eqref{def:E} and \eqref{def:K}, we obtain
\begin{equation}\label{eq:densK}
\rho^*=\frac{K_1(T^*,\mu^*;\bar{y}_{\rm max})}{K_0(T^*,\mu^*;\bar{y}_{\rm max})}.
\end{equation}

Since the maximum position $\bar{y}_{\rm max}$ satisfies the extremum equation \eqref{eq:bar_y}, the formula \eqref{eq:densK} simplifies to the remarkably simple form
\begin{equation}\label{rho_vs_T_mu}
	\rho^*=T^* \bar{y}_{\rm max}(T^*,\mu^*).
\end{equation}
This result provides an interesting direct relationship between the reduced density and the maximum position $\bar{y}_{\rm max}$.

In the literature on simple liquid theory \cite{HansenMcDonald13}, $\rho^*$ represents the reduced particle number density. Within our cell model framework, it has an additional interpretation as the average number of particles per cell, or the mean cell occupancy (cf. p.~\pageref{pRED} and~\cite[(2.26)]{KKD20}).

\subsection{Equation of state}

We have found the pressure $P^*$ and the particle density $\rho^*$, both as functions of thermodynamic variables $T^*$ and $\mu^*$, in equations \eqref{eos:reduced} and
\eqref{rho_vs_T_mu}, respectively.
To obtain the equation of state in the conventional form $P^* = P^*(T^*,\rho^*)$, we need to eliminate the chemical potential $\mu^*$ between these two relations.

Formally, we can solve the equation \eqref{rho_vs_T_mu} for $\mu^*$ to obtain $\mu^* = \mu^*(T^*,\rho^*)$, and then substitute it into \eqref{eos:reduced}. This procedure yields the pressure as a function of temperature and density (cf.~\cite[(2.28)]{KKD20}),
\begin{equation}
	\label{eos}
	P^*=T^* E\left[T^*,\mu^*(T^*,\rho^*); \frac{\rho^*}{T^*}\right],
\end{equation}
the equation of state in the widely used $P$-$\rho$-$T$ representation.

However, the inversion of equation \eqref{rho_vs_T_mu} to solve for $\mu^*(T^*,\rho^*)$ analytically is not feasible, and we must resort to numerical methods. Despite this limitation, the formal representation \eqref{eos} of equation of state proves valuable for theoretical analysis, particularly in determining critical point coordinates, as discussed in Section~\ref{sec:CPc}.

\section{Single-phase stability breakdown and critical points}\label{sec:CP}
In this section, we explore the breakdown of the single-phase regime established in Section~\ref{SLM}. As was concluded during the analysis of Fig.~\ref{fig:YvsMu_a1}, there exist some temperature range, below $T^*=0.4$ and above $T^*=0.2$, where the nature of $\bar{y}$ changes from a single-valued to a multi-valued function of $\mu^*$. A point $(T^*_c, \mu^*_c)$ where such change gradually appears is called a critical point. A stricter definition of the critical point will be given later in Section~\ref{sec:CPc}. 
The fact that the function $\bar{y}$ becomes multi-valued allows for the existence of multiple extrema for $E(T^*,\mu^*;y)$, and thus indicates the onset of phase coexistence in the system. 

The existence of critical points and emergence of phase coexistence can be understood by inspecting the behavior of $E(T^*,\mu^*;y)$ as a function of $y$.

\begin{figure}
	\includegraphics[width=1.0\textwidth,angle=0]{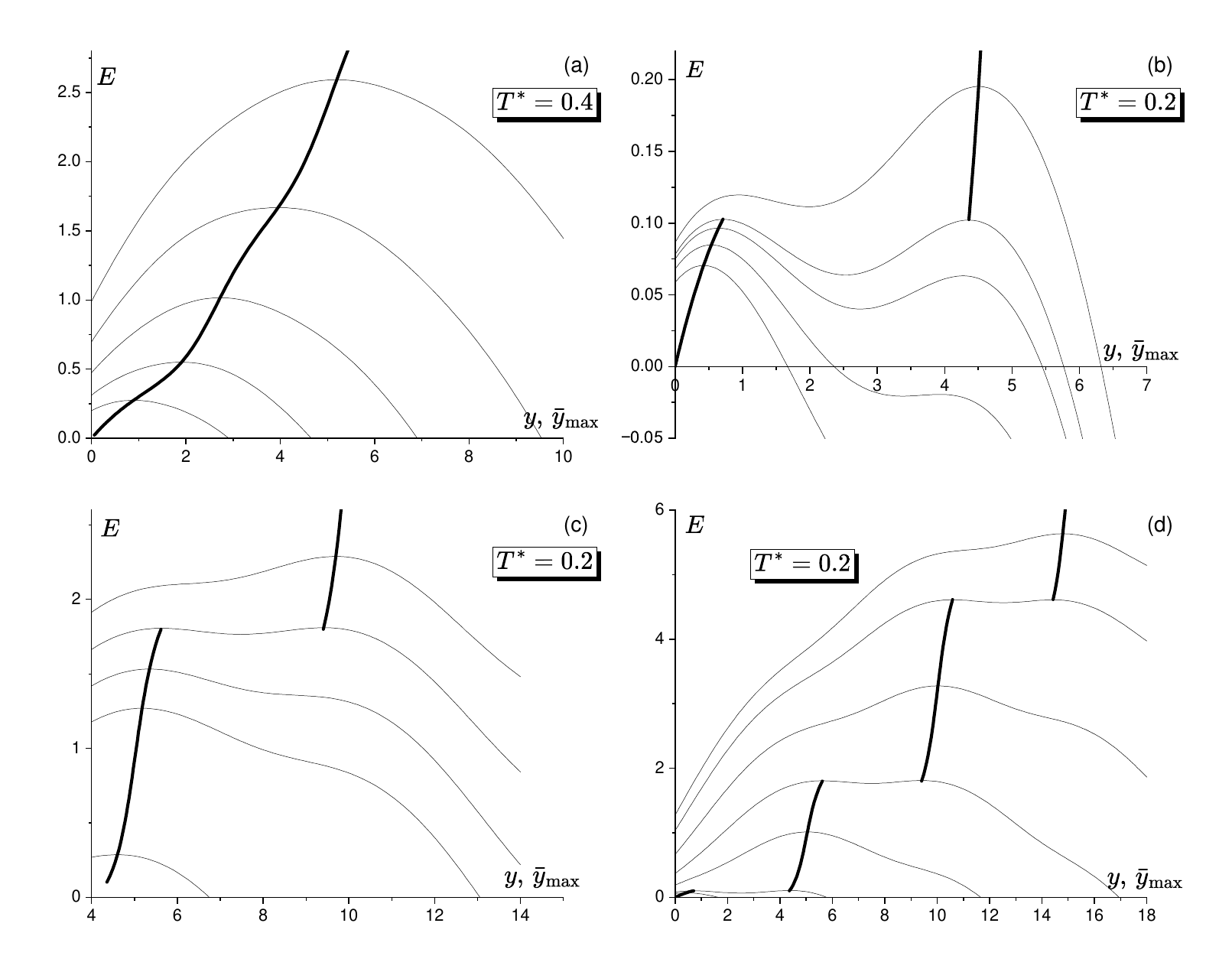}
\caption{\label{fig:EvsY}
Evolution of $E(T^*,\mu^*;y)$ with varying chemical potential.
Thin curves: $E(T^*,\mu^*;y)$ vs $y$ for specified values of $T^*$ and $\mu^*$.
		(a): $T^*=0.4$ (supercritical); $\mu^* = -0.1, \, 0.1, \, 0.3, \, 0.5, \, 0.7$.
		(b): $T^*=0.2$ (subcritical); $\mu^* = 0.2, \, 0.23, \, 0.25, \, 0.2592$
             (the two maxima have equal height), $0.28$.
		(c): $T^*=0.2$; $\mu^* = 0.3, \, 0.5, \, 0.55, \, 0.6, \, 0.65$.
		(d): $T^*=0.2$; $\mu^*$ varies from $0.2$ to $0.95$.
Plots are ordered bottom to top by increasing $\mu^*$.
Thick curves show the dependence of global maximum values
$E(T^*,\mu^*;\bar{y}_{\rm max})$ on $\bar{y}_{\rm max}$.
Auxiliary parameters $a$ and $v^*$ in $E(T^*,\mu^*;y)$
(see \eqref{def:E} and \eqref{def:K}) are fixed at $a=1.2$ and $v^*=5$.}
\end{figure}

Figure~\ref{fig:EvsY}(a) illustrates several plots of $E(T^*,\mu^*;y)$ versus $y$ at fixed temperature $T^*=0.4$ for different values of the chemical potential.
Each individual curve exhibits a single global maximum, which is characteristic of the single-phase domain where $T^*>T^*_c$.
The position of each maximum, $\bar{y}_{\rm max}$, corresponds to a particular pair of variables $T^*$ and $\mu^*$ .
The solid thick curve connects all these maxima, effectively tracing the maximum values $E(T^*,\mu^*;\bar{y}_{\rm max})$ as a function of $\bar{y}_{\rm max}$
at fixed temperature $T^*=0.4>T^*_c$ and varying $\mu^*$.
When viewed as a dependence of $E(T^*,\mu^*;\bar{y}_{\rm max})$ on $\mu^*$, the solid thick curve in Figure~\ref{fig:EvsY}(a) is a continuous function of $\mu^*$. This dependence has a counterpart in Figure~\ref{fig:YvsMu_a1}, where the argument $\bar{y}_{\rm max}(\mu^*)$ of $E(T^*,\mu^*;\bar{y}_{\rm max})$ is given by the single-valued, monotonically increasing dashed black curve $\bar{y}=\bar{y}_{\rm max}(\mu^*)$ plotted at the same $T^*=0.4$. The curves in both figures exhibit the same character of growth.

Figure~\ref{fig:EvsY}(b) shows several plots of $E(T^*,\mu^*;y)$ at $T^*=0.2<T^*_c$, illustrating its behavior below the critical temperature for different values of $\mu^*$.
For small chemical potential values ($\mu^*\lesssim0.23$), each curve exhibits only a single maximum at small $y$ values, similar to the supercritical case in Figure~\ref{fig:EvsY}(a).
However, as $\mu^*$ increases, the character of the function $E(T^*,\mu^*;y)$ changes dramatically. At $\mu^*\approx0.23$, an inflection point with a horizontal tangent appears, and with further increase of $\mu^*$, a second maximum develops at larger $y$ values, as illustrated by the curve with $\mu^*=0.25$.
The emergence of two maxima creates a local minimum between them, where the stability condition $E_2<0$ from \eqref{cond:max} is violated. As $\mu^*$ continues to increase, the relative heights of the two maxima evolve: the second maximum grows until it matches (at $\mu^*\approx0.2592$%
\footnote{All numerical calculations are performed with fixed values $a=1.2$ and $v^*=5$.})
and then exceeds the height of the first maximum.
At this point, the second maximum starts to dominate and thus determine
the thermodynamic behavior of the system.
Eventually, with sufficiently large $\mu^*$, the first maximum weakens, becoming an inflection point before disappearing entirely.
The solid thick curves trace the values $E(T^*,\mu^*;\bar{y}_{\rm max})$
of the dominant maxima at each chemical potential $\mu^*$, similarly as in the
Figure~\ref{fig:EvsY}(a).

If we continue to increase $\mu^*$, see Fig.~\ref{fig:EvsY}~(c), the third maximum appears, that eventually exceeds the second one and becomes the main one at $\bar{y}_{\rm max}\approx10$.
It seems that for the considered model such a process can be continued infinitely, and each time a new maximum will appear for larger and larger $y$. Figure~\ref{fig:EvsY}~(d), illustrates this idea by combining results of Figs.~\ref{fig:EvsY}~(b), and~\ref{fig:EvsY}~(c), and extending to higher values of $y$.

\label{pOSC}
This specific behavior can be understood by recognizing that the function $K_0(T^*,\mu^*; y)$ in \eqref{def:K} is equivalent to a deformed exponential function (see p.~\pageref{pDEF} and references quoted there). This function exhibits a periodic asymptotic behavior as $y\to\infty$, a phenomenon discovered by de Bruin \cite{DeBruijn49,DeBruijn53}, with oscillations manifesting at order $O(1)$. As $\mu^*$ increases, it progressively ``activates'' the maxima of these oscillations, revealing them one by one at larger and larger values of $y$. Graphical illustrations of the oscillating behavior of $K_0$ and related functions can be found in~\cite{DS24arxiv}, while explicit asymptotic formulas in terms of the Jacobi theta function $\vartheta_3$ are derived in~\cite{GS25}.

As in Figures~\ref{fig:EvsY}~(b)-(d), the solid blue curve for $\bar{y}=\bar{y}(\mu^*)$ in Figure~\ref{fig:YvsMu_a1} is plotted at the same temperature $T^*=0.2$.
Comparing it with the separate branches of $E(T^*,\mu^*;\bar{y}_{\rm max})$ in
Figures~\ref{fig:EvsY}~(b)-(d), we see that each of these branches corresponds to sections of the curve $\bar{y}=\bar{y}(\mu^*)$ with positive derivatives with respect to $\mu^*$.
The jumps in $\bar{y}(\mu^*)$ occur at those values of $\mu^*$ where new sections with positive slope start in the multi-valued function $\bar{y}(\mu^*)$ with increasing $\mu^*$.

In the present section, all graphical illustrations and observations are based on numerical computations. However, there is a limiting case --- the strong-repulsion limit $J_2\gg J_1$, or $a\gg1$ --- where all mathematics can be done explicitly, see
\cite[Sec. 7]{DS24arxiv}. In this special case, the sum over $n$ in the definition \eqref{def:K} of the function $K_0$ is limited to $n=1$. Consequently, only
two maxima can be observed, rather than an infinite number.
Nevertheless, the behavior near the origin is very similar. For example, Figure 6 in \cite{DS24arxiv} is qualitatively quite similar to Figure~\ref{fig:EvsY}(a), though the numerical input parameters are somewhat different.

At subcritical temperatures, there are intervals in $y$ where $E(T^*,\mu^*;y)$
possesses both maxima and minima.
Following the usual terminology of classical Landau theory, we can call the states corresponding to the highest, global, maxima {\it stable}, those corresponding to the lower local maxima --- {\it metastable}, and, finally, the ones corresponding to the local minima --- {\it unstable}.

The above analysis suggests that in the considered model there exist multiple first-order phase transitions at sufficiently low temperatures, and there are no phase transitions at sufficiently high temperatures. The question arises whether those first-order phase transitions end with corresponding critical points? The most natural way to investigate this question in many-body interacting systems is in terms of pressure and density. We will proceed with such an investigation.

\section{Equation of state and critical point coordinates}\label{sec:CPc}

General thermodynamic theory of phase transitions provides us with a recipe for finding the coordinates of a critical point from an equation of state in the form $P=P(T,\rho)$,
see e.g.~\cite[p.~184]{HansenMcDonald13}.
Since the critical point is a horizontal inflection point of the critical isotherm
at the top of the phase coexistence curve \cite[Fig.~5.12]{HansenMcDonald13}, we have two conditions:
\begin{equation}\label{split}
		\left(\frac{\partial P^*}{\partial \rho^*}\right)_{T^*} = 0
		\qquad\mbox{and}\qquad
		\left(\frac{\partial^2 P^*}{\partial \rho^{*2}}\right)_{T^*} = 0
\end{equation}
to be satisfied at the critical point $(T^*_c,\rho^*_c)$. The physical content of the first condition in \eqref{split} is that the isothermal compressibility diverges at the
critical point.

In Appendix \ref{CPC}, we establish the connection between the standard
critical point conditions \eqref{split} and the corresponding equations specific for the system we consider throughout the paper. These special equations are
$E_2(T^*,\mu^*;\bar{y}_{\rm max})=0$ and $E_3(T^*,\mu^*;\bar{y}_{\rm max})=0$, where
the functions $E_2$ and $E_3$ are the derivatives of the basic function
$E(T^*,\mu^*;y)$, see \eqref{def:E}, \eqref{Ej}, evaluated at the stable-maximum position $y=\bar{y}_{\rm max}$.

Along with equation \eqref{cond:max} determining the maximum's position
$\bar{y}_{\rm max}$, we have three equations in terms of functions $E_j$ and variables $T^*$, $\mu^*$, and $\bar{y}_{\rm max}$:
\begin{equation}\label{eq:system}
	\begin{cases}
		E_1(T^*,\mu^*; \bar{y}_{\rm max}) = 0,\\
		E_2(T^*,\mu^*; \bar{y}_{\rm max}) = 0,\\
		E_3(T^*,\mu^*; \bar{y}_{\rm max}) = 0.
	\end{cases}
\end{equation}
This system of three equations in three unknowns can be solved numerically to determine the critical point coordinates $T^*_c$, $\mu^*_c$, and $\bar{y}_{\rm max, c}$.

In order to ensure that the coordinate $\bar{y}_{\rm max, c}$ indeed corresponds to a global maximum of $E(T^*,\mu^*;y)$, we still have to perform a negativity check (similar to that in \eqref{cond:max})
\begin{equation}\label{poss}
E_4(T^*_c,\mu^*_c; \bar{y}_{\rm max, c}) < 0
\end{equation}
distinguishing it from other possible stationary points.

A note on the application of the Laplace's method at the critical point is in place here.
Since $E_2(T^*,\mu^*; \bar{y}_{\rm max})=0$ at the critical point, the usual Laplace approximation formula~\eqref{laplace1} is not applicable in this case.
However, the inequality \eqref{poss} allows us to use a modified version of the Laplace's method with the result \cite[6.4.19d]{BenderOrszag99}
\begin{equation}
	\label{eq:Xi_4}
	\Xi = \sqrt{\frac{N_v T^*}{2\pi}} \frac{2\Gamma(1/4) (4!)^{1/4}}{4[-N_v E_4(T^*_c,\mu^*_c)]^{1/4}} \exp[N_v E(T^*_c,\mu^*_c;\bar{y}_{\rm max, c})],
\end{equation}
where $\Gamma(\cdot)$ is the Euler Gamma function.
In the thermodynamic limit $N_v \to \infty$, the formula \eqref{eq:Xi_4} leads to the same expression for pressure as given in \eqref{eos:reduced}.

\section{Multiple critical points in the cell model}\label{MCP}
The cell model of fluid defined in Section \ref{sec:model} is known to possess multiple critical points~\cite{KKD18,KKD20,KD22} associated with cascades of first-order phase transitions.
The system of equations \eqref{eq:system} defining the critical point coordinates has infinitely many solutions. This special feature can be attributed to the oscillatory periodic behavior of functions $E_j$ appearing in \eqref{eq:system}, see p.~\pageref{pOSC}.
As in the above references, we label such solutions by natural numbers $n$, starting with $n=1$ for the lowest critical temperature.

\begin{table}[t]
	\centering
	\caption{Critical point coordinates and thermodynamic quantities for the first five critical points ($n=1,\ldots,5$) of the cell model. The solutions are obtained numerically using Maple for the system of equations~\eqref{eq:system} with parameters $a=1.2$ and $v^*=5$.}
	\begin{tabular}{c|c|c|c|c|c|c}
		\toprule
		$n$ & $T^*_c$ & $\mu^*_c$ & $\bar{y}_c$ & $\beta^*_c$ & $\rho^*_c$ & $P^*_c$ \\
		\midrule
		1 & 0.254567 & 0.209380 & 2.01870 & 3.92823 & 0.513896 & 0.0503397  \\
		2 & 0.261881 & 0.585097 & 5.74906 & 3.81852 & 1.50557 & 0.437695  \\
		3 & 0.265254 & 0.891843 & 9.43632 & 3.76996 & 2.50303 & 1.05817 \\
		4 & 0.267242 & 1.16893  & 13.1039 & 3.74193 & 3.50191 & 1.89369 \\
		5 & 0.268562 & 1.42932  & 16.7608 & 3.72354 & 4.50131 & 2.93781 \\
		\midrule
		$n$ & $E(T^*_c,\mu^*_c)$ & $E_{1,c}$; $E_{2,c}$; $E_{3,c}$ & $E_4(T^*_c,\mu^*_c)$ & $\frac{T_c(n)}{T_c(n-1)}$ & $\frac{T_c(n)}{T_c(n=1)}$ & $S^*_c$ \\
		\midrule
		1 & 0.197746 & 0.0 & -0.120306 & $-$ & 1.00000 & 2.43173 \\
		2 & 1.67135  & 0.0  & -0.113509 & 1.02873 & 1.02873 & 1.34933 \\
		3 & 3.98927  & 0.0  & -0.110425 & 1.01288 & 1.04198 & 0.914901 \\
		4 & 7.08603  & 0.0 & -0.108631 & 1.00749 & 1.04979 & 0.631157 \\
		5 & 10.9390  & 0.0 & -0.107455 & 1.00494 & 1.05497 & 0.417427 \\
		\bottomrule
	\end{tabular}
	\label{tab:cp}
\end{table}

Table~\ref{tab:cp} presents the numerical solutions for the first five critical points from the infinite sequence, along with the corresponding values of density, pressure, and other thermodynamic quantities evaluated at the critical point coordinates $T^*_c$, $\mu^*_c$, and $\bar{y}_c$.

Similarly to previous studies~\cite{KKD18,KKD20,KD22,DKPP25}, we have chosen $a=1.2$ for the interaction strength ratio $a$ (denoted by $f$ in \cite{KD22}) in the numerical calculations. Our choice $v^*=5$ for the reduced cell volume can be useful in looking for similarities in numerical data from different papers. Including the column for $\beta^*_c(n)=1/T^*_c(n)$ in Table~\ref{tab:cp} facilitates comparison as this value coincides with $p_c^{(n)}$
used in the above references. Table~\ref{tab:cp} extends consideration of critical points from $n=1,2,3$ to $n=1,\ldots,5$.

For example, the numerical value of $\beta^*_c(1)=1/T^*_c(1)$ is the same as in \cite[p.~ 12]{KKD20} and applies to the data displayed in \cite[Table 2]{KD22}.
Moreover, $\beta^*_c(n)$ agree with corresponding entries of \cite[Table~2]{DKPP25} for $n=1,\,2$, and $3$.
Similarly, our numerical values of $\rho^*_c$ for $n=1,2,3$ are in full agreement with that for $\bar n_c^{(n)}$ from \cite[Table~2]{DKPP25}, and the data in \cite[Table~1]{KKD20} and \cite[Table~2]{DKPP25} computed at different values of $v$.
These observations suggest certain universality of reduced temperatures and particle densities calculated in different settings (recall that the integration variable $z=y+\beta\mu+\ln v$ is used in \cite{KD22} instead of $y$ as in \cite{KKD20} and the current paper).
Remarkably, the reduced density $\rho^*_c$ shows an approximately linear relationship with the critical point index $n$, with $\rho^*_c \approx n-1/2$ for the whole range of $n$ considered.

Another interesting observation is that critical temperatures $T^*_c(n)$ exhibit a monotonic increase with $n$, with decreasing increments. The ratio $T^*_c(n)/T^*_c(n-1)$ decreases from $1.02873$ for $n=2$ to $1.00494$ for $n=5$, suggesting convergence toward a limiting temperature, say
$$
T^*_c(\infty)=\lim_{n\to\infty}T^*_c(n)\,.
$$
It is natural to expect that the temperature $T^*_c(\infty)$ is just a reciprocal of the marginal value $p_0$ introduced in \cite[Theorem~2.1]{KKD20}: If the temperature exceeds $T^*_c(\infty)=p_0^{-1}$, the system cannot undergo any phase transitions anymore, and is in a single-phase state.

We expect that the limiting temperature $T^*_c(\infty)$ is a smooth function of the interaction ratio $a>1$.
Moreover, in the strong-repulsion limit $a\gg1$, it is expected that $T^*_c(\infty)$ tends to the value $1/4$ in agreement with $p_c(10)=4.0000$ computed in \cite[Table~1]{KKD20} (see also the last two rows in \cite[Table~2]{DKPP25}) and $p_c(a\to\infty)=4$ found analytically in \cite[Sec.~7.1]{DS24arxiv}.

\begin{figure}[htbp]
	\includegraphics[width=0.5\textwidth,angle=0]{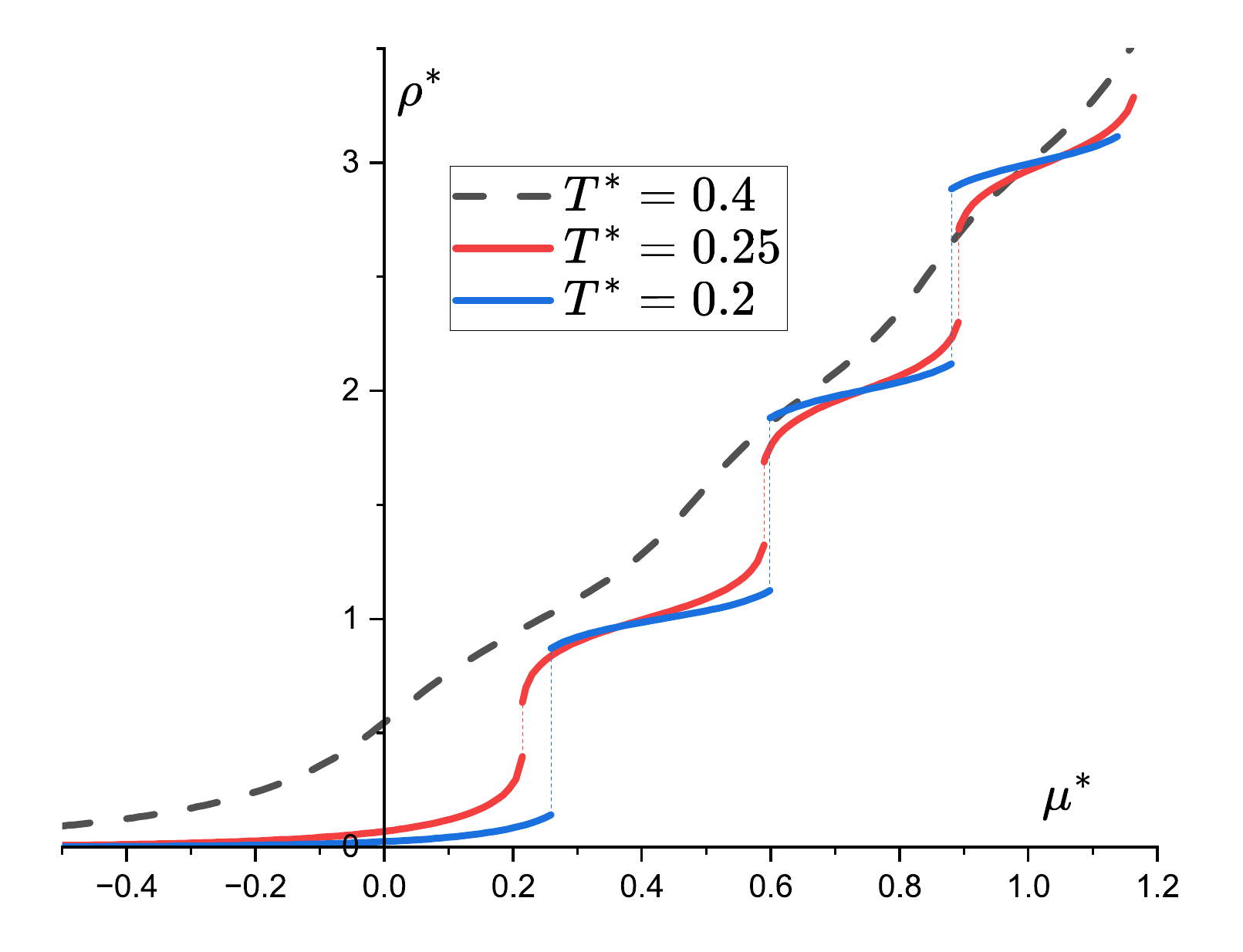}
	\centering
	\captionsetup{width=0.9\textwidth}
	\caption{The particle number density $\rho^*(T^*,\mu^*)$ as a function of $\mu^*$ at different values of temperature: Dashed black curve: $T^*=0.4$; Red: $T^*=0.25$; Blue: $T^*=0.20$. Thin lines denote corresponding two-phase regions on the corresponding isotherms.}
	\label{fig:rho_vs_mu_a1}
\end{figure}

\begin{figure}
	\includegraphics[width=0.49\textwidth,angle=0]{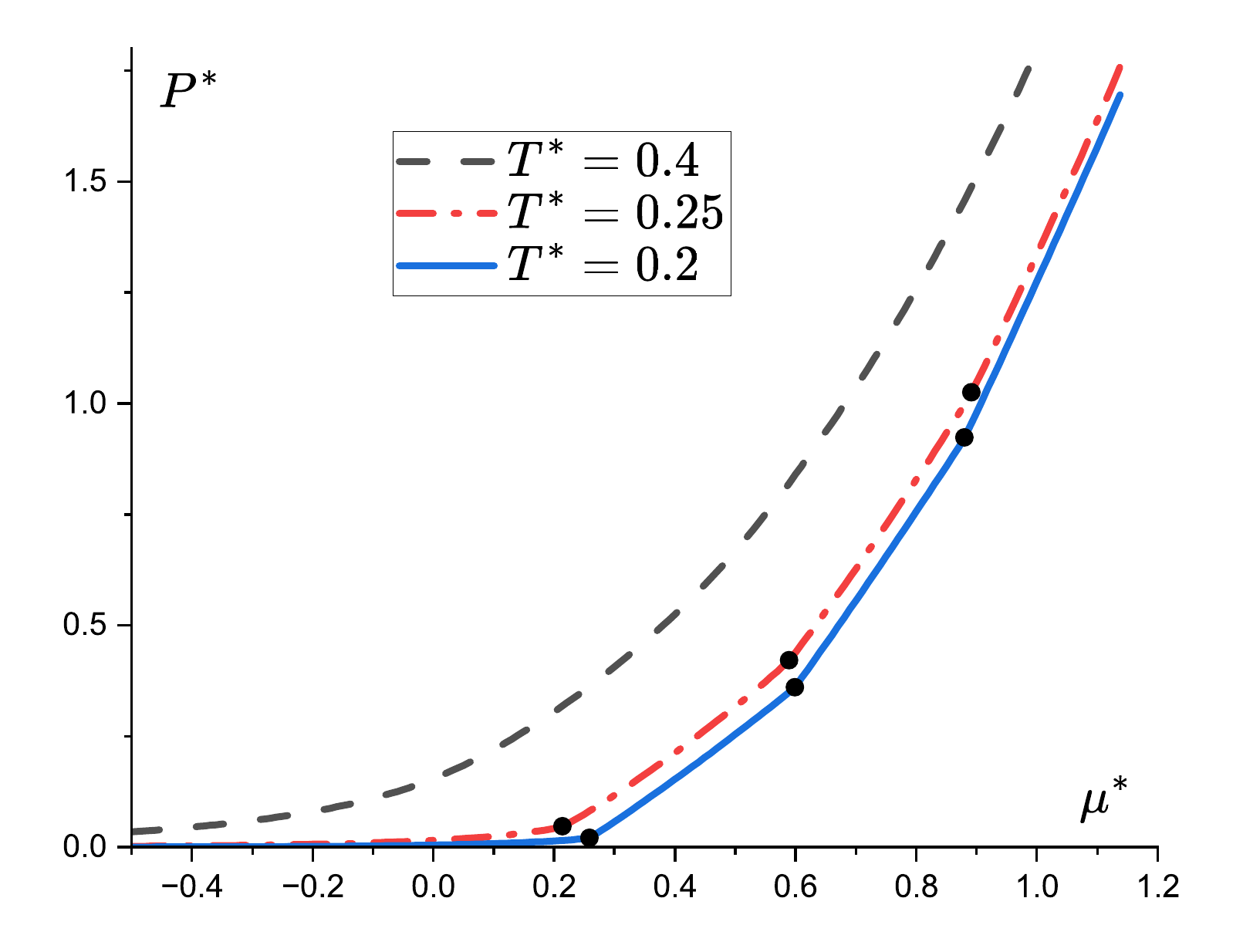}
	\hfill
	\includegraphics[width=0.49\textwidth,angle=0]{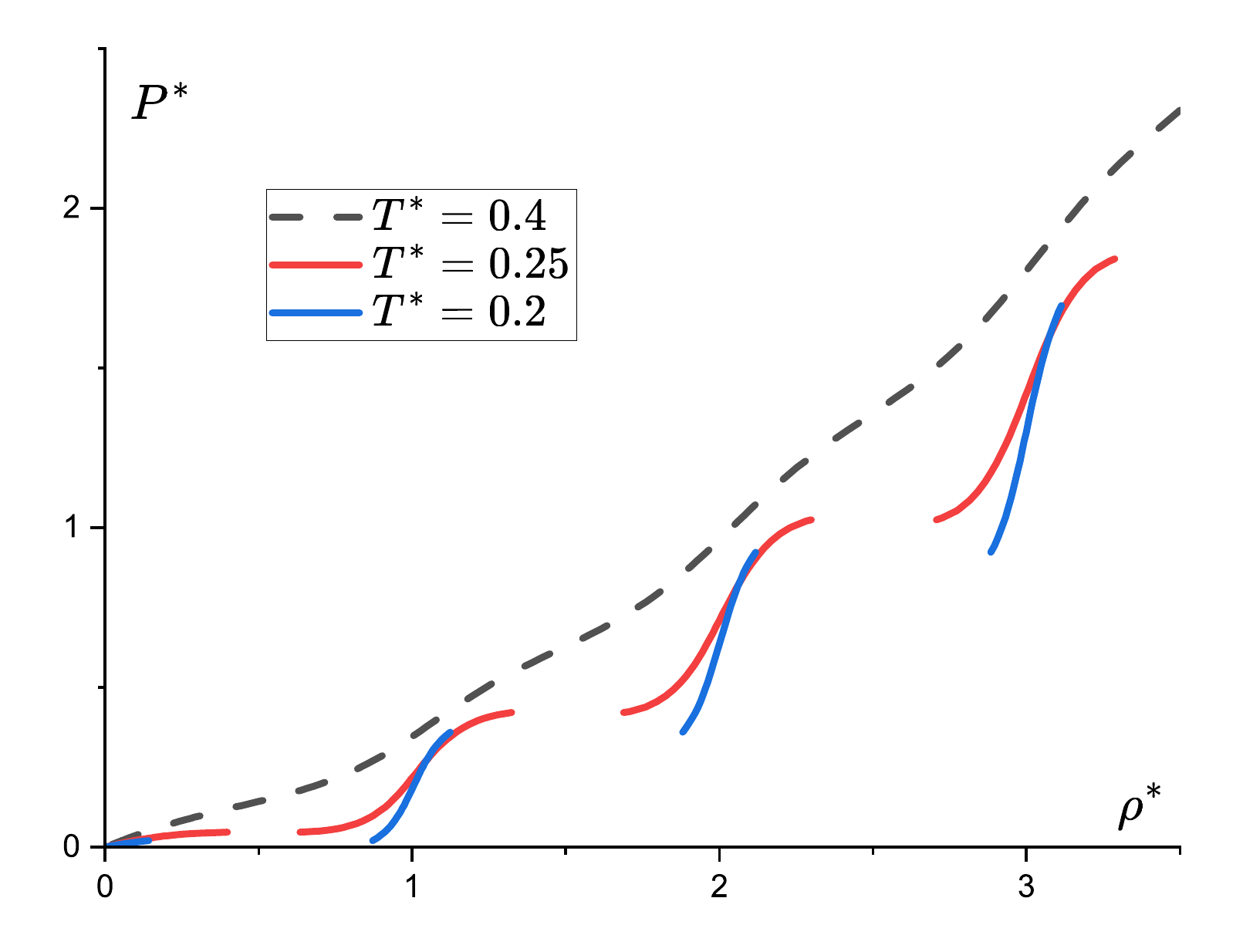}
	\\
	\parbox{0.45\textwidth}{\caption{\label{fig:p_vs_mu_a1} Pressure $P^*(T^*,\mu^*)$ as a function of $\mu^*$ at different values of temperature: Dashed curve (black): $T^*=0.4$; Dash-dotted curve (red): $T^*=0.25$; Solid curve (blue): $T^*=0.20$. Full circles denote the points of phase transitions.}}
	\hfill
	\parbox{0.45\textwidth}{\caption{\label{fig:p_vs_rho_a1}Pressure $P^*$ as a function of $\rho^*$ at different values of temperature: Dashed black curve: $T^*=0.4$; Red: $T^*=0.25$; Blue: $T^*=0.20$.}}
\end{figure}

\subsection{Graphical illustrations}
Taking into account results of Section~\ref{sec:CP}, we can now illustrate some results from previous sections graphically.

The dependence of the number density $\rho^*$ on the chemical potential $\mu^*$ at three representative values of temperature $T^*$ is illustrated in Fig.~\ref{fig:rho_vs_mu_a1}.
The corresponding temperature values are (i) $T=0.4$, which is greater than any critical temperature, (ii) $T=0.25$, which is slightly lower than the critical temperature values, and (iii) $T=0.2$, which is much lower than the critical temperatures.

The dependence of the pressure $P^*$ on the chemical potential $\mu^*$ at the same three temperatures is demonstrated in Fig.~\ref{fig:p_vs_mu_a1}.

The pressure as a function of density is shown in Fig.~\ref{fig:p_vs_rho_a1}. Note that the subcritical isotherms intersect. This density anomaly has been observed for a few interaction potentials, for example, for a ramplike discretized potential~\cite{NBBS06}, and for a continuous-shouldered-well potential~\cite{VF10}. We plan to investigate implications of this anomaly in our future works.

\section{Entropy}
\label{sec:entropy}
By thermodynamic formulas for the grand canonical ensemble, we have for the entropy
\begin{equation}
	S = -\left(\frac{\partial \Omega}{\partial T}\right)_{V,\mu} = V\left(\frac{\partial P}{\partial T}\right)_{\mu}.
\end{equation}

At this point, there are two possibilities to normalize entropy. One possibility is to normalize it by the average number of particles
\begin{equation}
	S^{*} = \frac{S}{k_{\rm B} \langle N \rangle},
\end{equation}
and call this quantity the entropy per particle. Another possibility is to normalize it by the number of cells
\begin{equation}
	S^{*}_v = \frac{S}{k_{\rm B}N_v},
\end{equation}
and call this quantity the entropy per cell. The usage of the entropy per particle, $S^*$, is more common in the theory of many-particle systems~\cite{HansenMcDonald13}, while the entropy per cell, $S^{*}_v$, may be useful in cell (or lattice) models. We will also call $S^*$ the \textit{reduced entropy}.

\textbf{Entropy per particle} reads
\begin{eqnarray}
	S^* & = & \frac{1}{\rho^*}\left(\frac{\partial P^*}{\partial T^*}\right)_{{\mu^*}}
	= \frac{(\partial P^* / \partial T^*)_{{\mu^*}}}{(\partial P^* / \partial \mu^*)_{T^*}}.
\end{eqnarray}
By~\eqref{eos:reduced}
\begin{equation}
	\label{eq:entropy}
	S^*(T^*,\mu^*) = \frac{1}{\rho^*(T^*,\mu^*)}
	\left[
	E(T^*,\mu^*;\bar{y}_{\rm max}) + T^* \left(\frac{\partial E(T^*,\mu^*;\bar{y}_{\rm max})}{\partial T^*}\right)_{\mu^*}
	\right].
\end{equation}
The result of calculation is
\begin{equation}
	\label{S_vs_T_mu}
	S^*(T^*,\mu^*) = \left(\frac{3}{2} - \frac{\mu^*}{T^*}\right) - \frac{1}{T^*}\frac{K_1}{K_0} + \frac{K_0 \ln K_0}{K_1} + \frac{a}{2T^*} \frac{K_2}{K_1},
\end{equation}
where for the special functions we imply $K_i = K_i(T^*,\mu^*;\bar{y}_{\rm max})$.
The first contribution to the entropy corresponds to the entropy of non-interacting molecules in a lattice, or an ideal lattice gas contribution, see~\cite[(47.4)]{Hill56}, if we account for the ideal-gas chemical potential $\mu_{\rm id} = k_{\rm B}T \ln(N\Lambda^3/V)$.

The entropy per cell $S^{*}_v$ is related to the entropy per particle $S^*$ via
\begin{equation}
	S^* = \frac{1}{\rho^*} S^{*}_v,
\end{equation}
where
\begin{equation}
	S^{*}_v  = \left(\frac{\partial P^*}{\partial T^*}\right)_{\mu^*}.
\end{equation}
Thus, the expressions for the entropy per cell is
\begin{eqnarray}
	\label{eq:entropy2}
	S^{*}_v(T^*,\mu^*) & = & \left(\frac{3}{2} - \frac{\mu^*}{T^*}\right)\frac{K_1}{K_0} - \frac{1}{T^*}\frac{K_1^2}{K_0^2} + \ln K_0 + \frac{a}{2T^*} \frac{K_2}{K_0}
\end{eqnarray}
Details of calculation are present in Appendix~\ref{sec:app:entropy}.

\begin{figure}[htbp]
	\includegraphics[width=0.5\textwidth,angle=0]{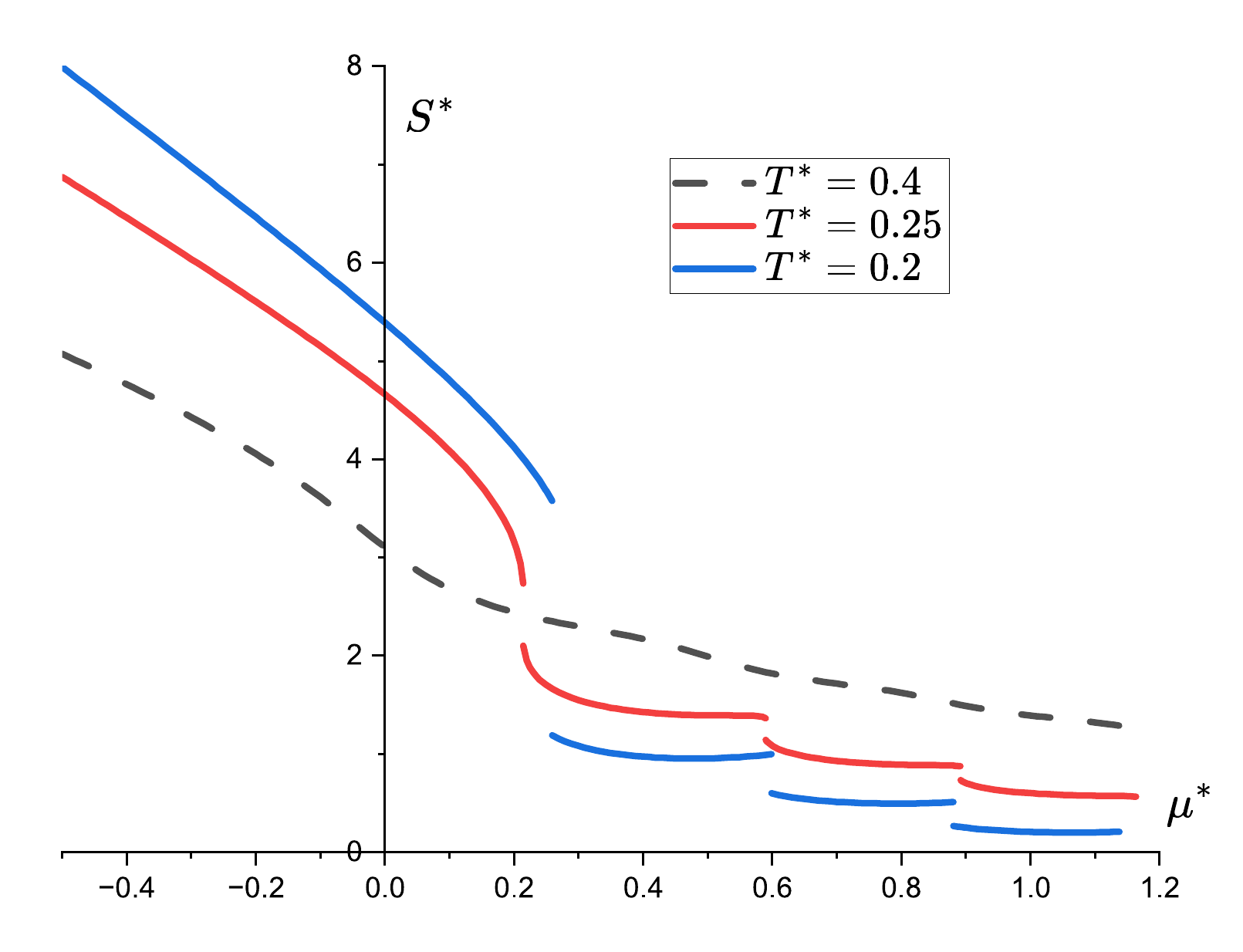}
	\hfill
	\includegraphics[width=0.5\textwidth,angle=0]{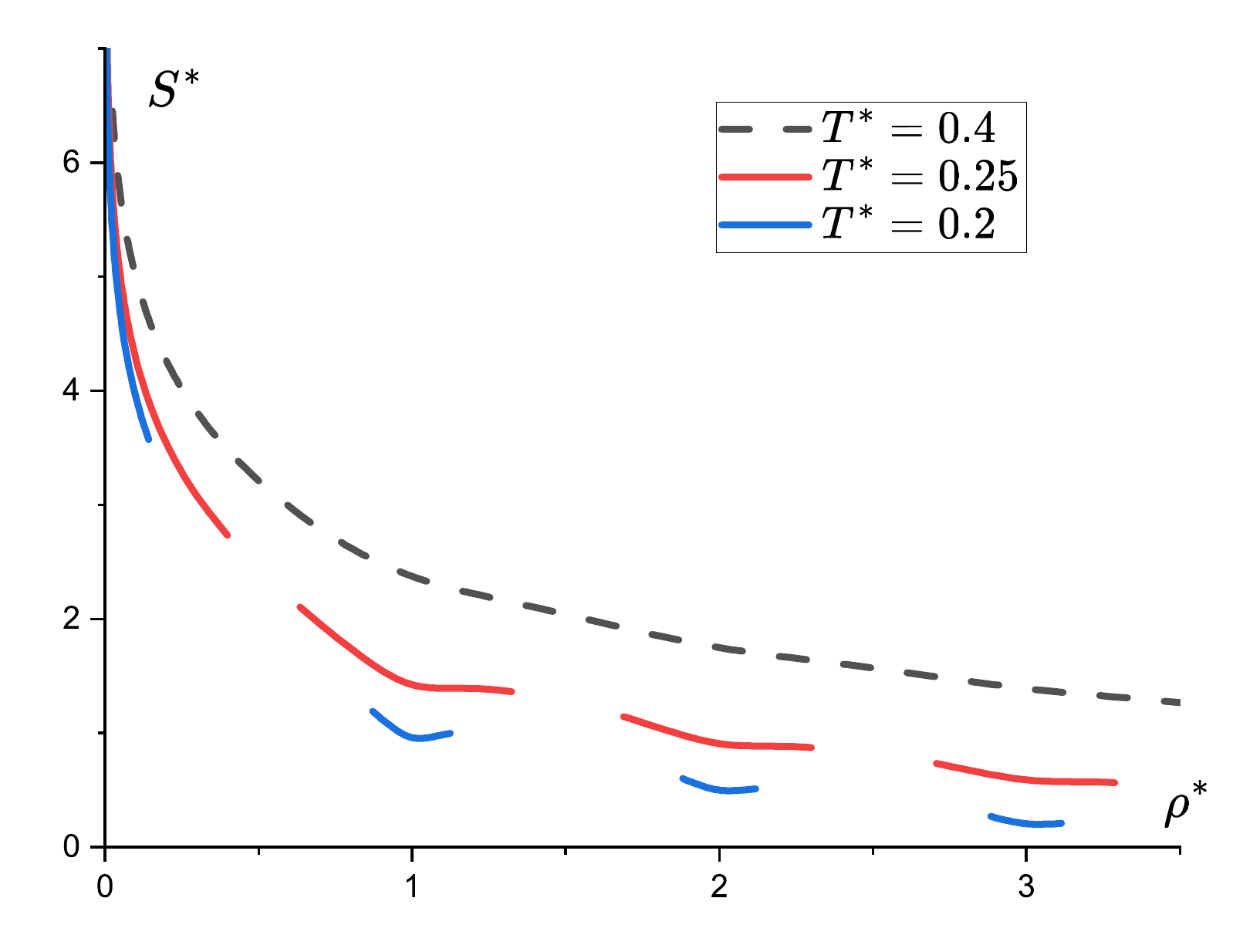}
	\\
	\parbox{0.48\textwidth}{\caption{\label{fig:S_vs_mu_a1} Entropy per particle $S^{*}$ as a function of $\mu^*$. Dashed black curve: $T^*=0.4$; Red: $T^*=0.25$; Blue: $T^*=0.2$.}}
	\hfill
	\parbox{0.48\textwidth}{\caption{\label{fig:S_vs_rho_a1} Entropy per particle $S^{*}$ as a function of $\rho^*$. Dashed black curve: $T^*=0.4$; Red: $T^*=0.25$; Blue: $T^*=0.2$.}}
\end{figure}

The dependence of $S^*$ on the chemical potential $\mu^*$ at some temperatures is presented in Fig.~\ref{fig:S_vs_mu_a1}.

The equation~\eqref{S_vs_T_mu} expresses the reduced entropy as a function of temperature and chemical potential. To obtain the reduced entropy as a function of temperature and density, we combine equations~\eqref{S_vs_T_mu} and~\eqref{rho_vs_T_mu} into a parametric equation, with $\mu^*$ being the parameter. This way we can plot the dependence of $S^*$ on density at a given temperature. Such dependencies are present in Fig.~\ref{fig:S_vs_rho_a1}.

A qualitative behavior of the entropy as a function of density, and as a function of chemical potential can be compared with analogous results for the double-occupancy Hubbard model considered in~\cite{DBKGS11}. In particular, in the plot for the entropy per particle in \cite[Fig.~8]{DBKGS11} we distinguish two regions of interest -- small densities ($\rho^* < 0.5$) and densities around 1 ($\rho^*\approx 1$). For small densities we observe that the entropy per particle is rapidly decreasing from high values at $\rho \approx 0$ to moderate values at $\rho^*\approx 0.5$. At densities $\rho^*\approx 1$ a typical `cusp-like' behavior is observed in both our results (Fig.~\ref{fig:S_vs_rho_a1}) and \cite[Fig.~8]{DBKGS11}. At sufficiently low temperatures, the entropy reaches a minimum at $\rho \approx 1$, then increase little bit. The investigation and analysis of the details of the entropy dependencies on temperature, chemical potential, and density deserves a dedicated research.

\section{Conclusion}
\label{sec:conclusion}
In this study, a many-particle interacting system is studied based on a cell model with unrestricted cell occupancy. The interaction between particles is defined through two contributions, namely, a global attractive one between all particle pairs, and a short-range repulsive one for particles within the same cell. It is shown that such model posesses multiple critical points. 
For low temperatures, the system exhibits a sequence of phase transitions of the first order between multiple phases, while for high temperatures there exists only a single phase.

The statistical-mechanical theory is derived within the framework of the grand canonical ensemble, and asymptotically exact solution~\eqref{laplace1} is obtained. The previously obtained results are revisited with explicit account for the de Briglie wavelength $\Lambda$ in the definition of the grand partition function, and, as a consequence, for the explicit temperature dependence of computed physical quantities, such as pressure, density, and entropy.
The values of these, and a few other, physical quantities are calculated at first five critical points and summarized in Table~\ref{tab:cp}.

The entropy of the considered model has been calculated for the first time. Explicit expressions are obtained for both the entropy per particle~\eqref{S_vs_T_mu} and entropy per cell~\eqref{eq:entropy2}. The dependence of the entropy per particle on the chemical potential, as well as that on the density is presented graphically in Fig.~\ref{fig:S_vs_mu_a1} and Fig.~\ref{fig:S_vs_rho_a1}, respectively. Qualitative similarity in behavior of the entropy is noticed between the considered model and the three-dimensional Hubbard model applied to cold atoms subject to an optical lattice~\cite{DBKGS11}. The obtained entropy results are of interest in general, and require a dedicated research.

Another non-trivial result is that the isotherms in the pressure-density plane intersect, see Fig.~\ref{fig:p_vs_rho_a1}. In this regard, we want to bring to our attention studies~\cite{VF10} and~\cite{NBBS06}, where such density anomaly was also found. A model with continuous-shouldered-well potential was studied in~\cite{VF10}, and that with a ramplike discretized potential in~\cite{NBBS06}, both papers discussing a possibility of more that one critical point in such systems.
To study of our model in this context is also in our plans.



\bmhead{Acknowledgements} This work was supported by the National Research Foundation of Ukraine under the project No. 2023.03/0201.

\bmhead{Data Availability Statement} Data sharing is not applicable to this article as no data sets were generated or analyzed during the current study.

\bmhead{Conflict of Interest Statement} The authors have no conflicts of interest.

\pagebreak
\begin{appendices}

\section{Figures}
This Appendix contains Figures demonstrating typical behavior for quantities $E$ \eqref{def:E}, $E_1$ \eqref{def:reducedE1}, and $E_2$ \eqref{def:reducedE2}, at different values of temperature $T^*$ and chemical potential $\mu^*$.

\subsection{Figures for $E(T^*,\mu^*;y)$}
\label{sec:figE0}

\begin{figure}[htbp]
	\includegraphics[width=0.3\textwidth,angle=0]{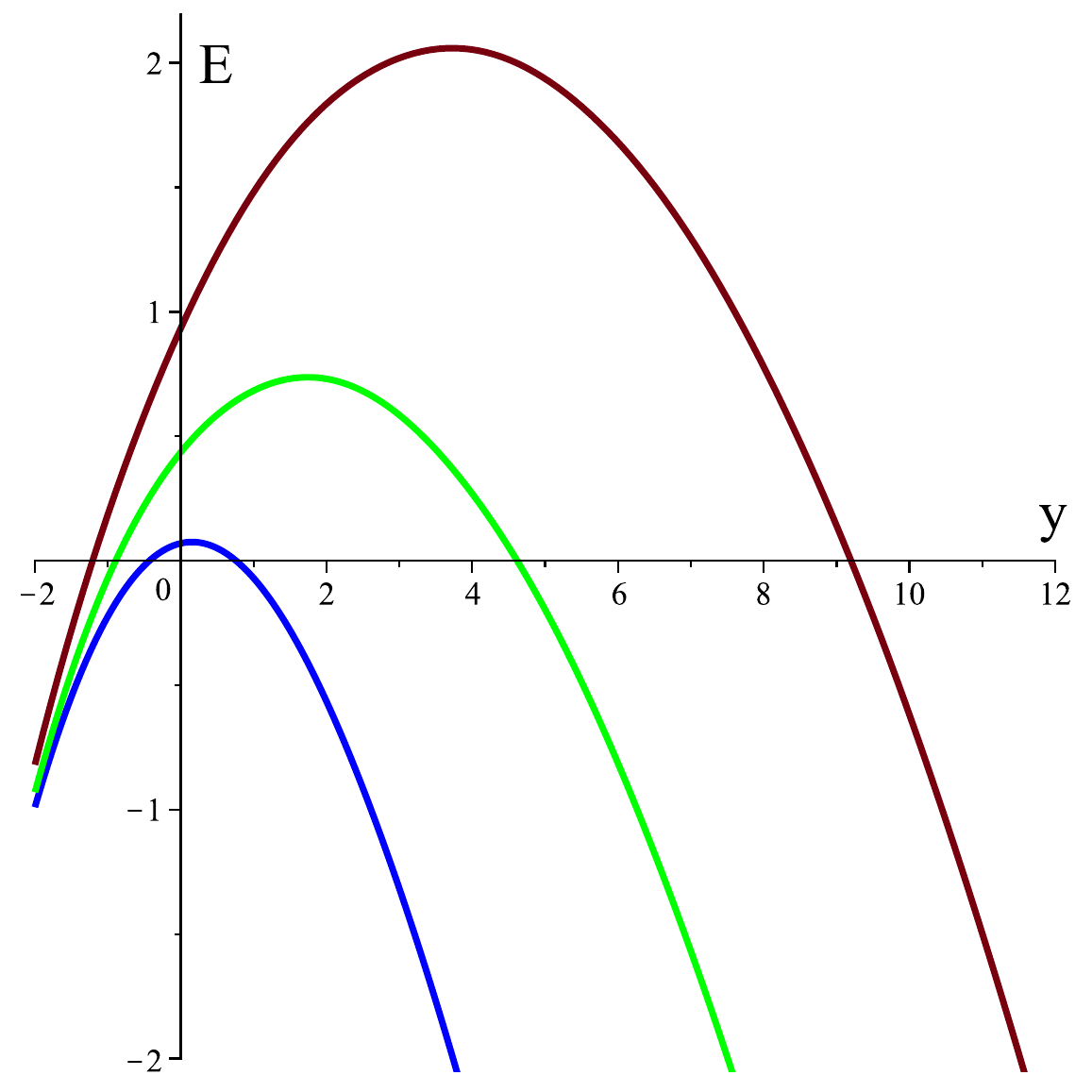}
	\hfill
	\includegraphics[width=0.3\textwidth,angle=0]{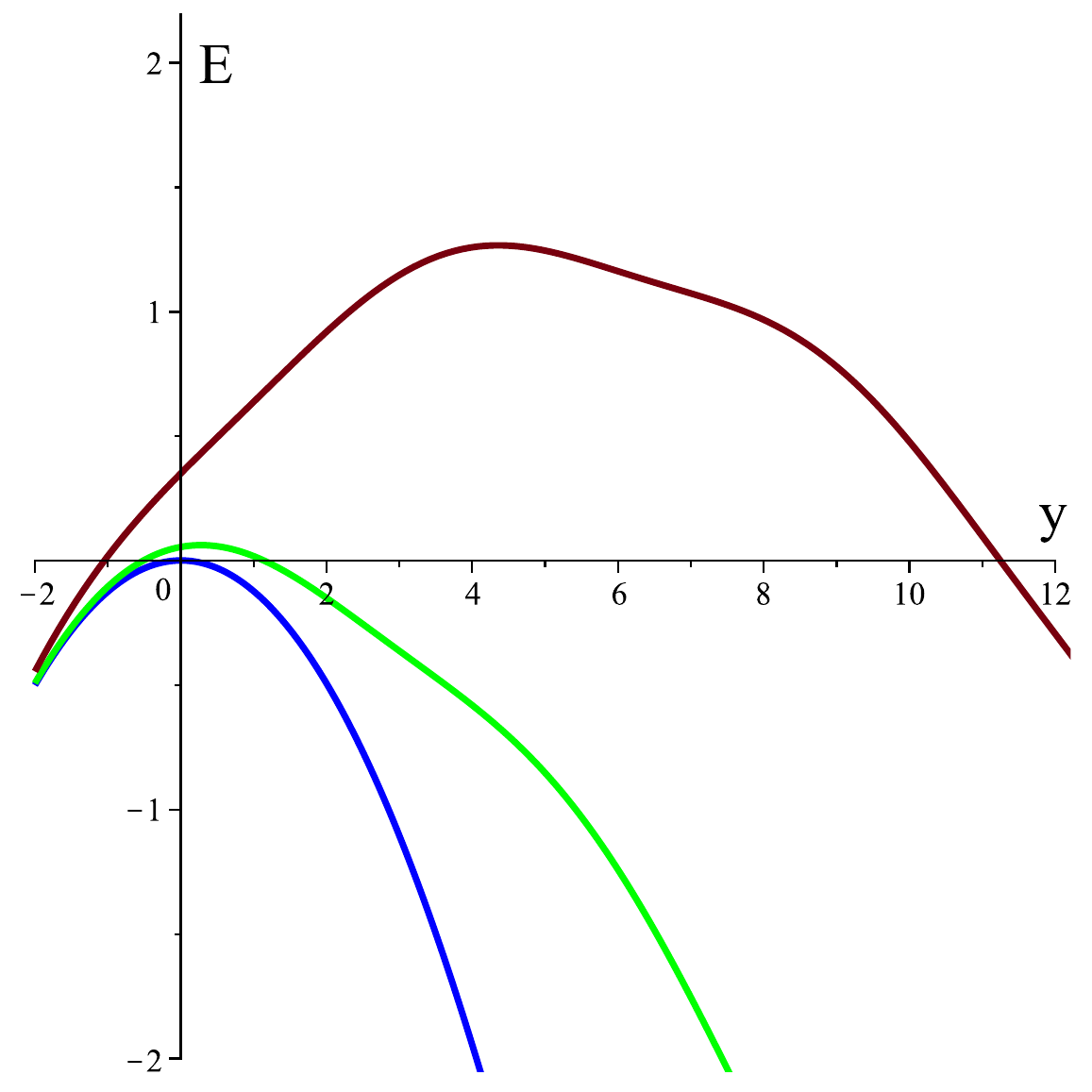}
	\hfill
	\includegraphics[width=0.3\textwidth,angle=0]{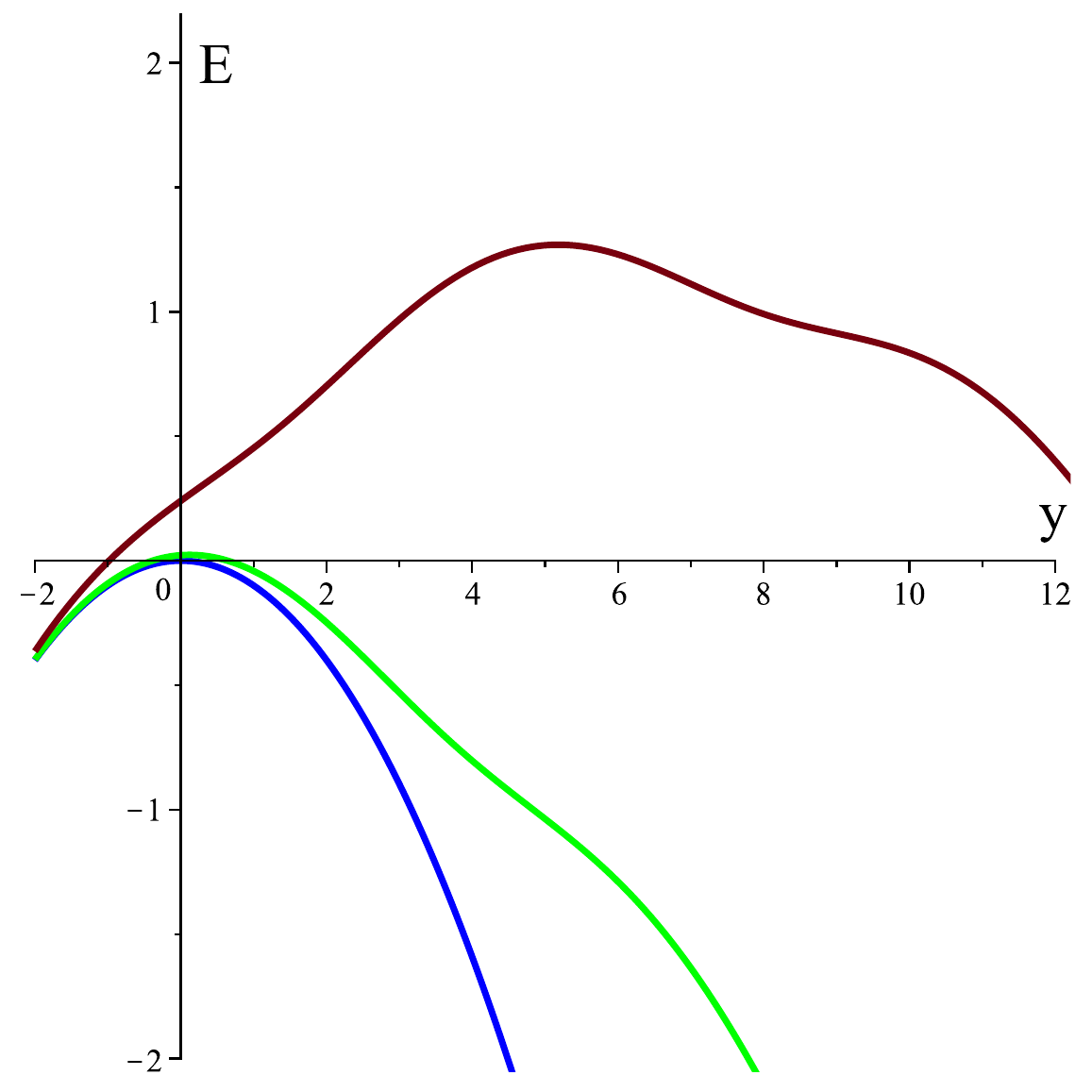}
	\\
	\vfill
	\parbox{0.95\textwidth}{\caption{\label{fig:E0y_vs_y_a} Quantity $E(T^*,\mu^*;y)$ as a function of $y$ at different values of $T^*$ and $\mu^*$. Left: $T^*=0.5$; Center: $T^*=0.25$; Right: $T^*=0.2$. Curves: Blue: $\mu^*=-1.0$; Green: $\mu^*=0.0$; Red: $\mu^*=0.5$.}}
	
\end{figure}

\begin{figure}[htbp]
	\includegraphics[width=0.3\textwidth,angle=0]{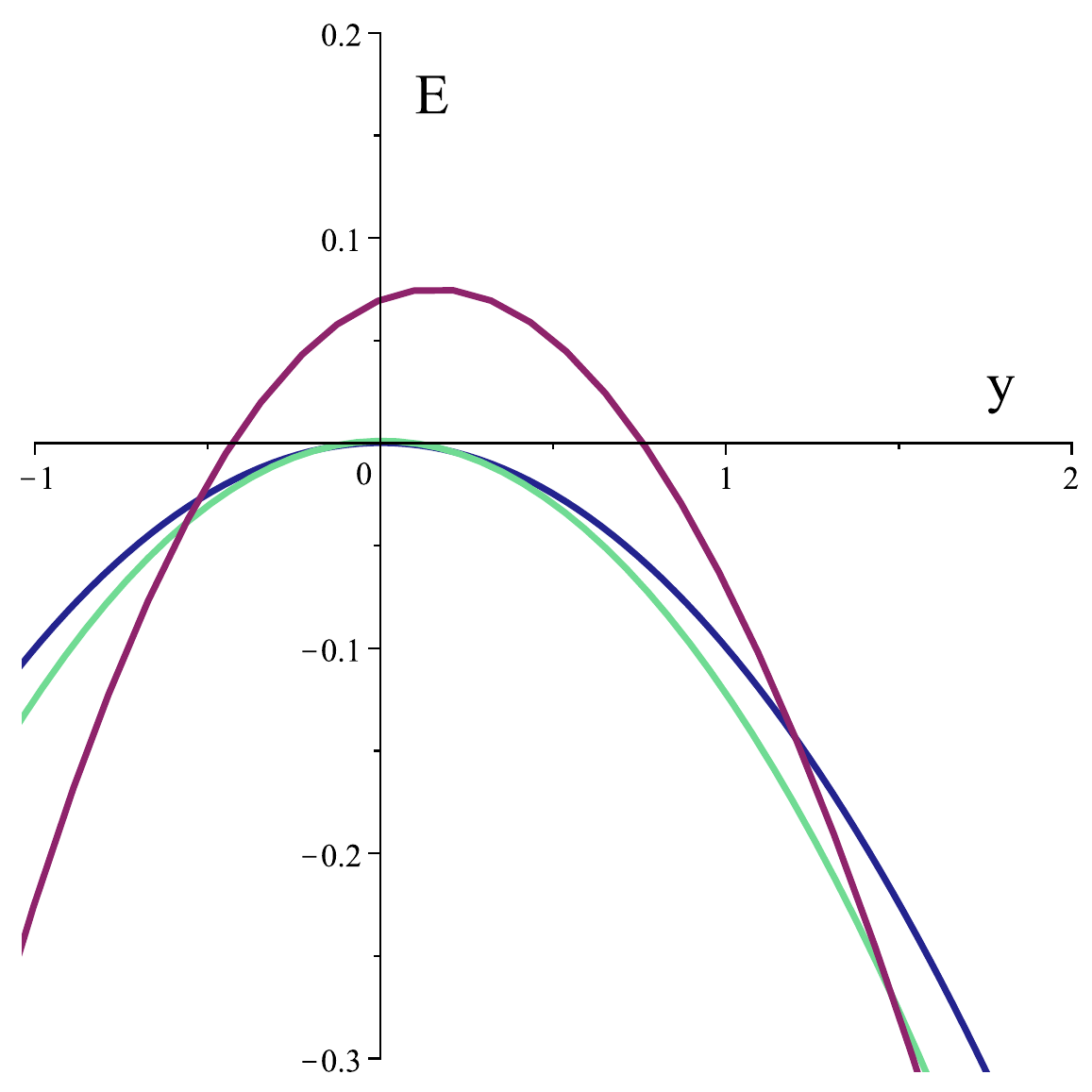}
	\hfill
	\includegraphics[width=0.3\textwidth,angle=0]{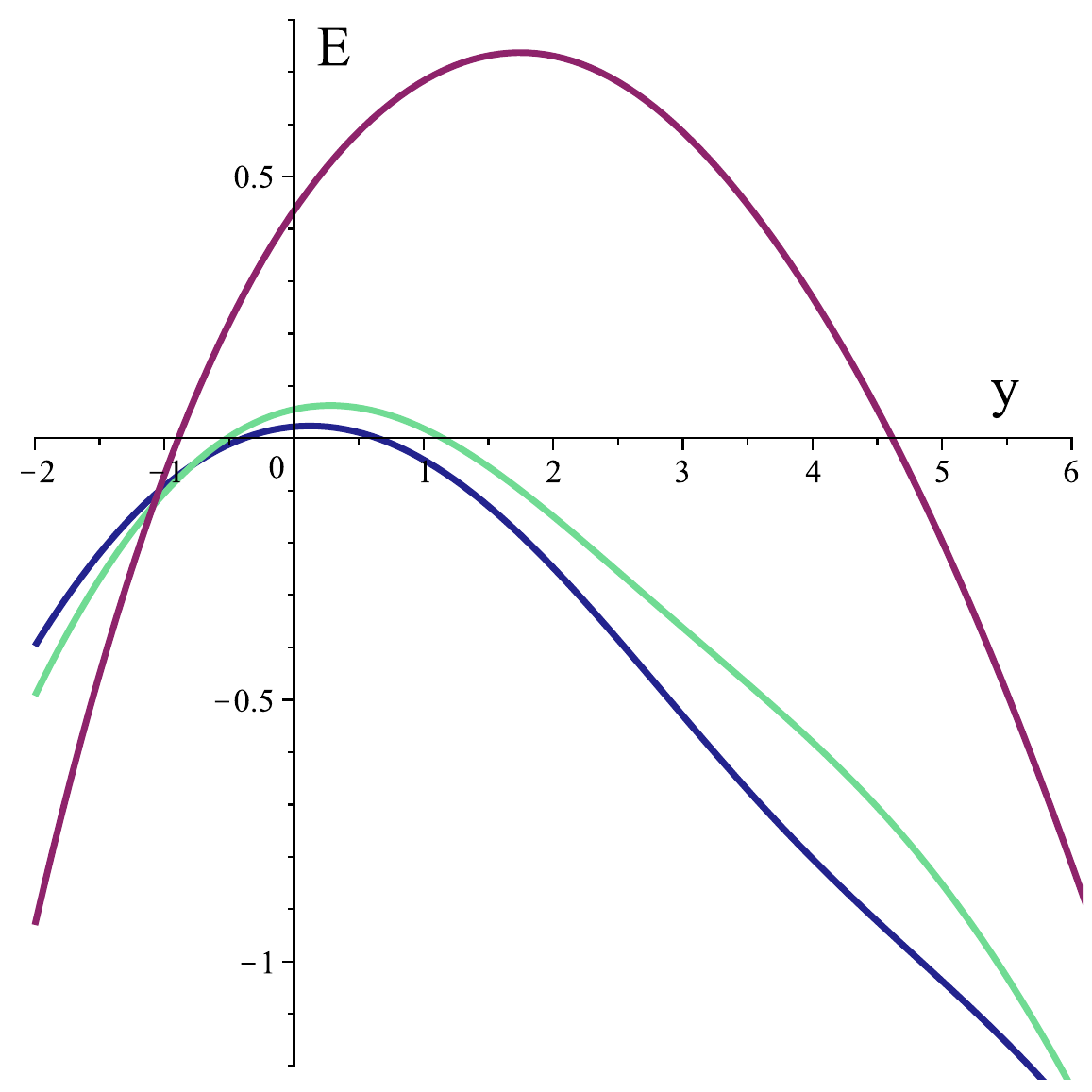}
	\hfill
	\includegraphics[width=0.3\textwidth,angle=0]{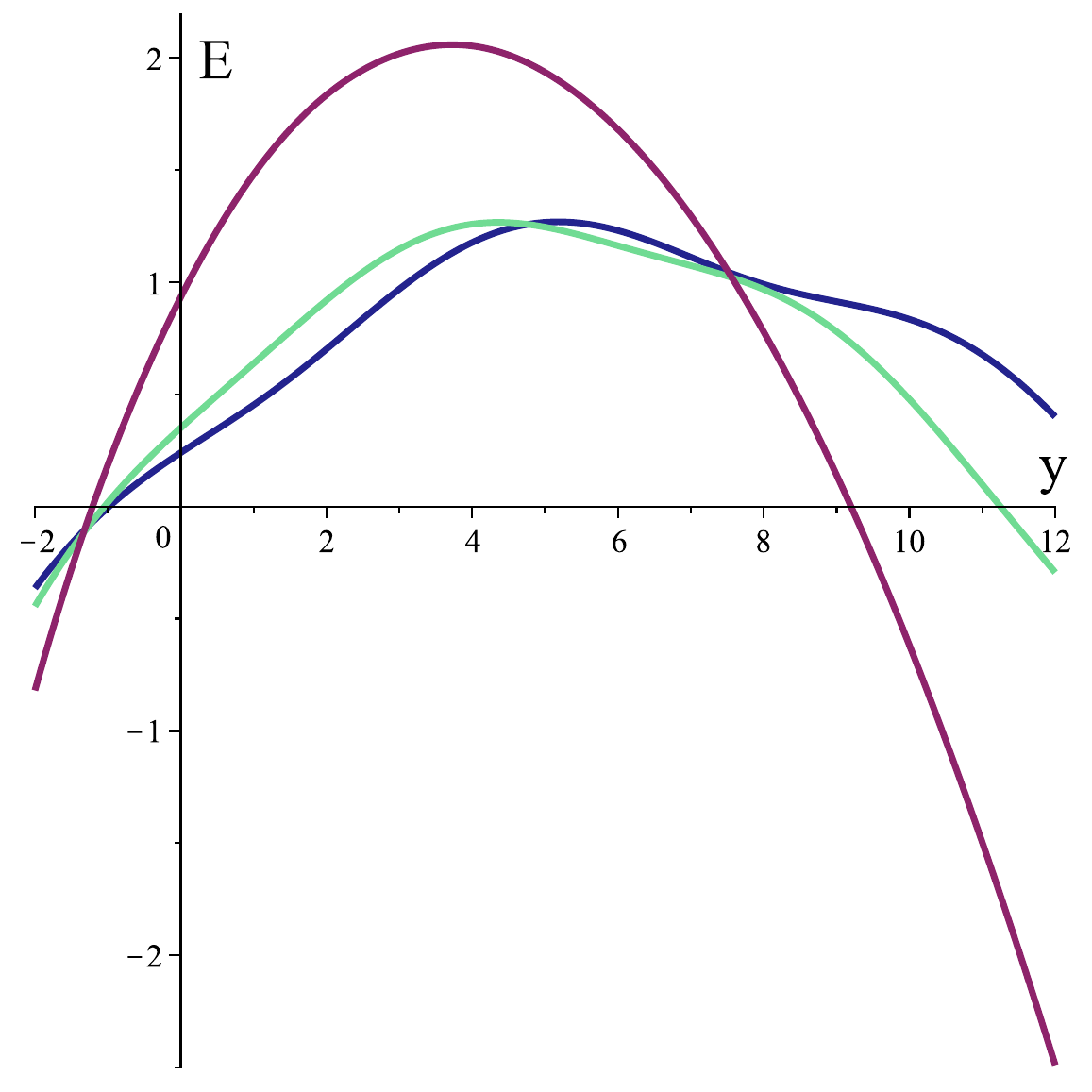}
	\vfill
	\parbox{0.95\textwidth}{\caption{\label{fig:E0y_vs_y_b} Quantity $E(T^*,\mu^*;y)$ as a function of $y$ at different values of $T^*$ and $\mu^*$. Left: $\mu^*=-1.0$; Center: $\mu^*=0.0$; Right: $\mu^*=0.5$. Curves: Navy: $T^*=0.2$; Aquamarine: $T^*=0.25$; Maroon: $T^*=0.5$.}}
	
\end{figure}

\pagebreak
\subsection{Figures for $E_1(T^*,\mu^*;y)$}
\label{sec:figE1}

\begin{figure}[htbp]
	\includegraphics[width=0.3\textwidth,angle=0]{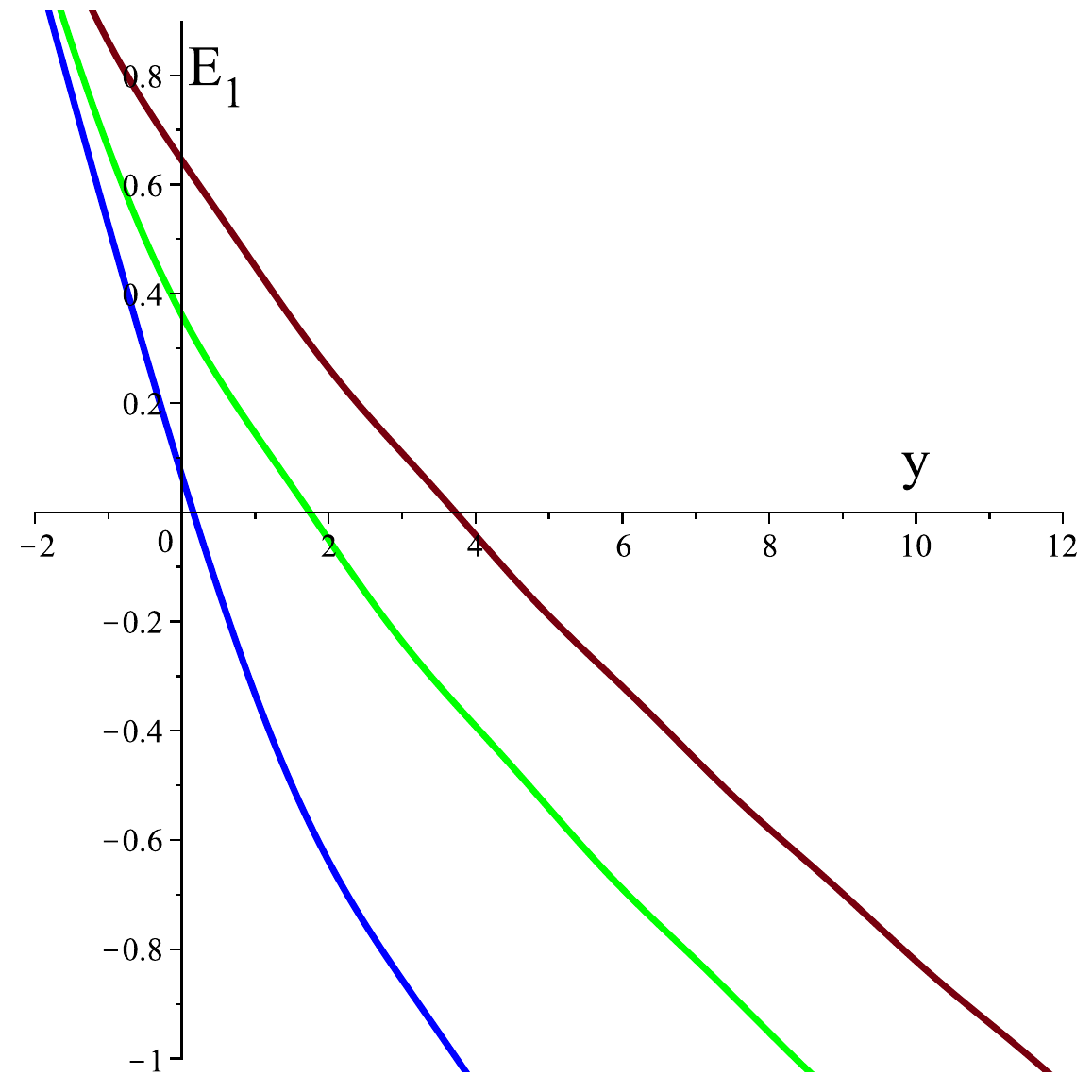}
	\hfill
	\includegraphics[width=0.3\textwidth,angle=0]{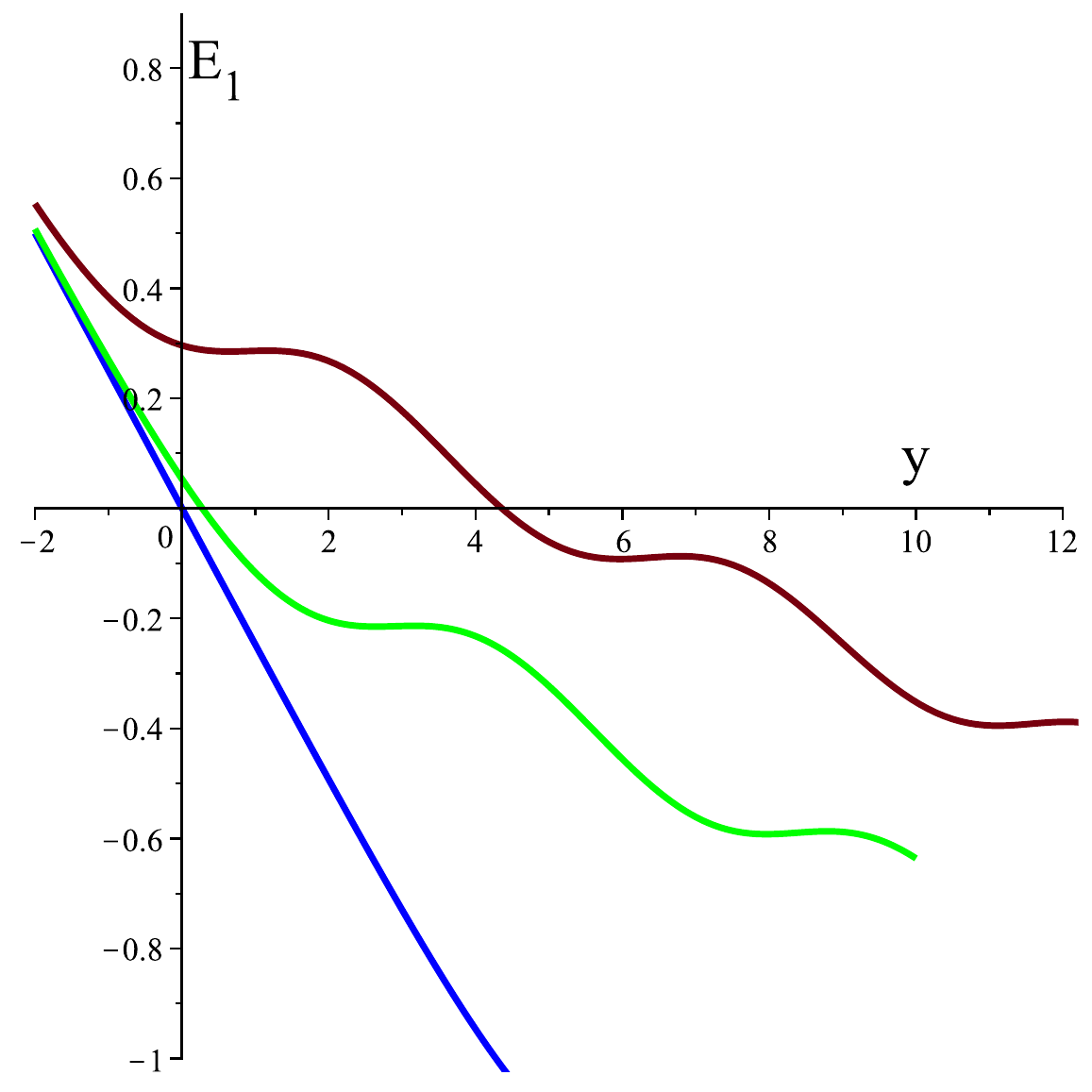}
	\hfill
	\includegraphics[width=0.3\textwidth,angle=0]{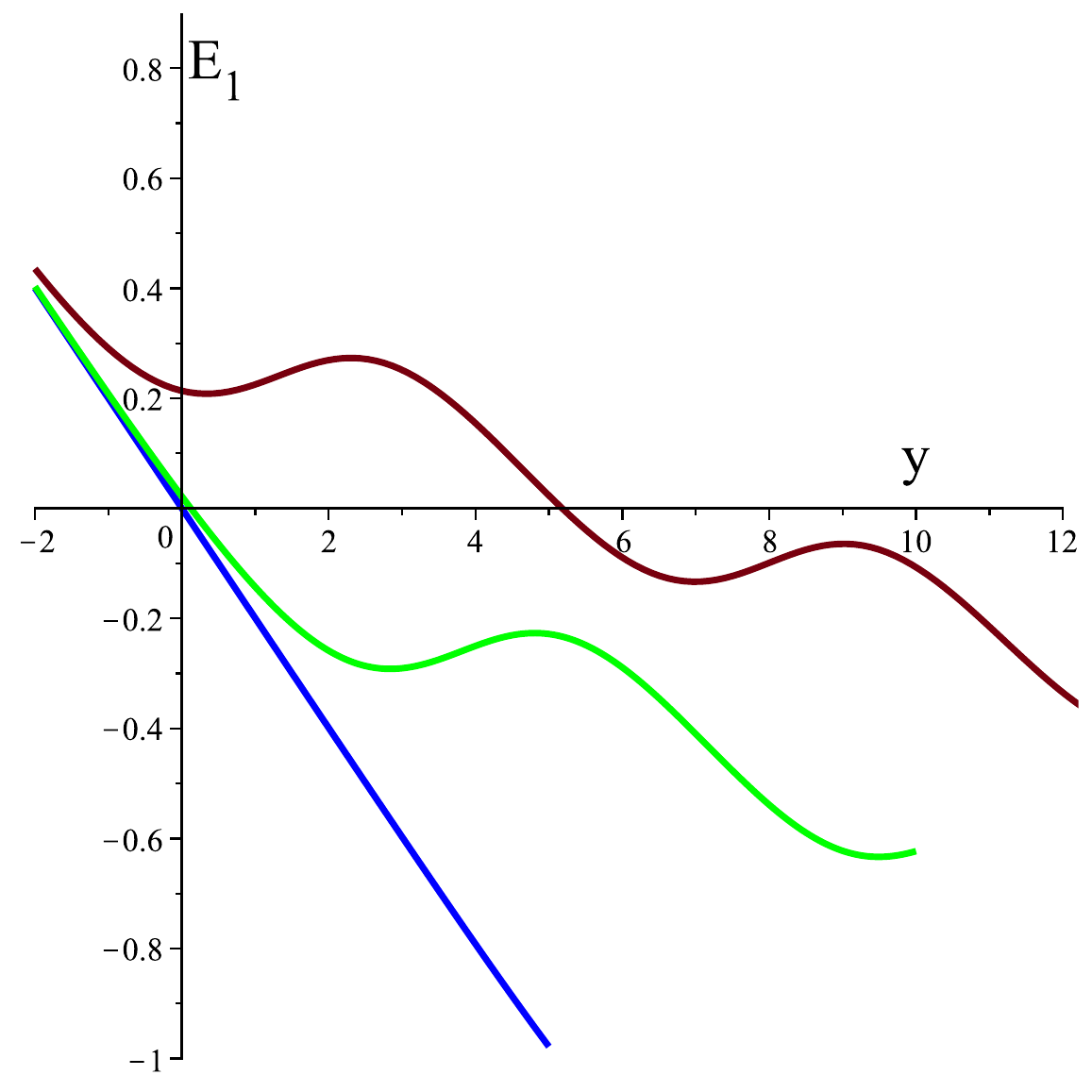}
	\\
	\vfill
	\parbox{0.95\textwidth}{\caption{\label{fig:E1y_vs_y_a} Quantity $E_1(T^*,\mu^*;y)$ as a function of $y$ at different values of $T^*$ and $\mu^*$. Left: $T^*=0.5$; Center: $T^*=0.25$; Right: $T^*=0.2$. Curves: Blue: $\mu^*=-1.0$; Green: $\mu^*=0.0$; Red: $\mu^*=0.5$.}}
	
\end{figure}

\begin{figure}[htbp]
	\includegraphics[width=0.3\textwidth,angle=0]{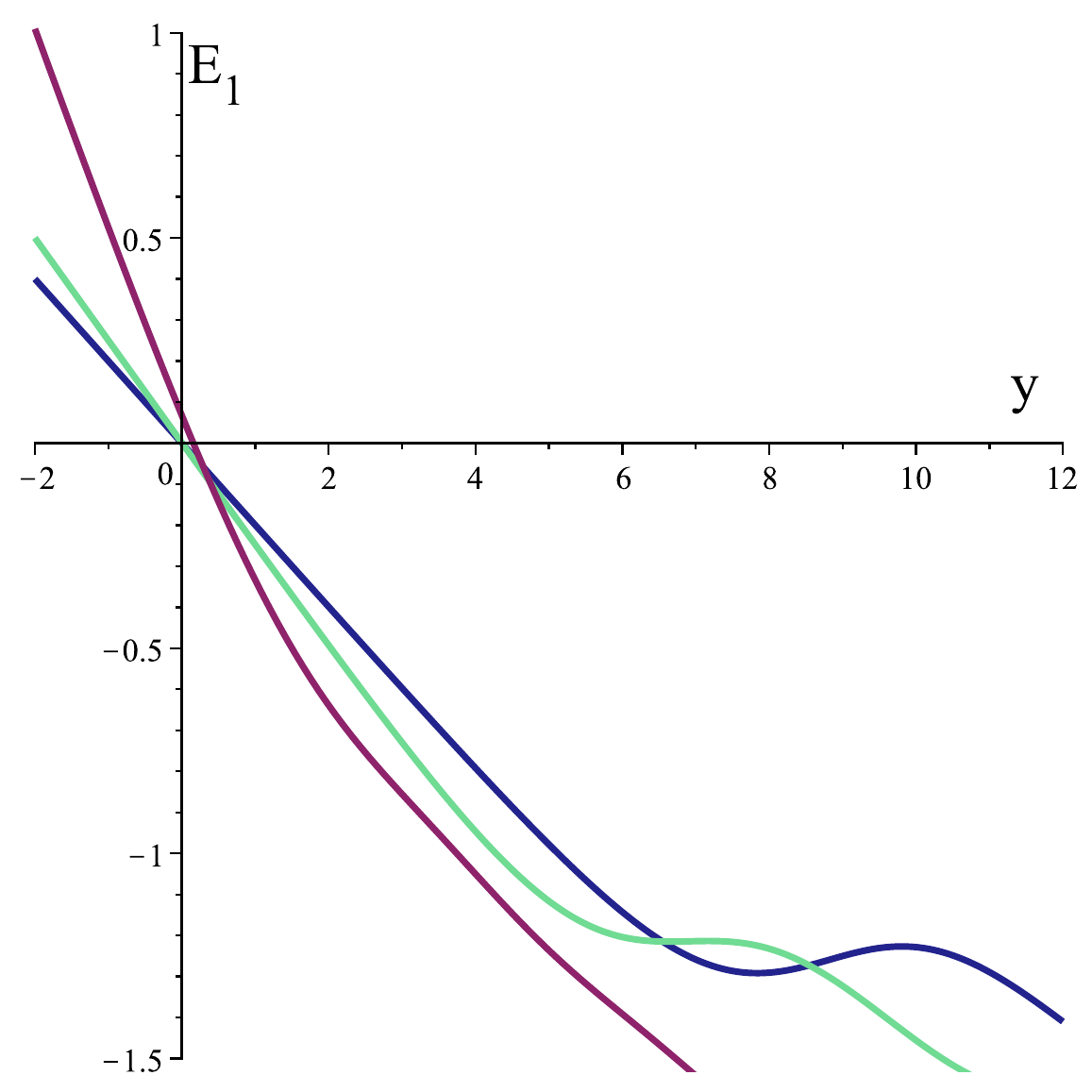}
	\hfill
	\includegraphics[width=0.3\textwidth,angle=0]{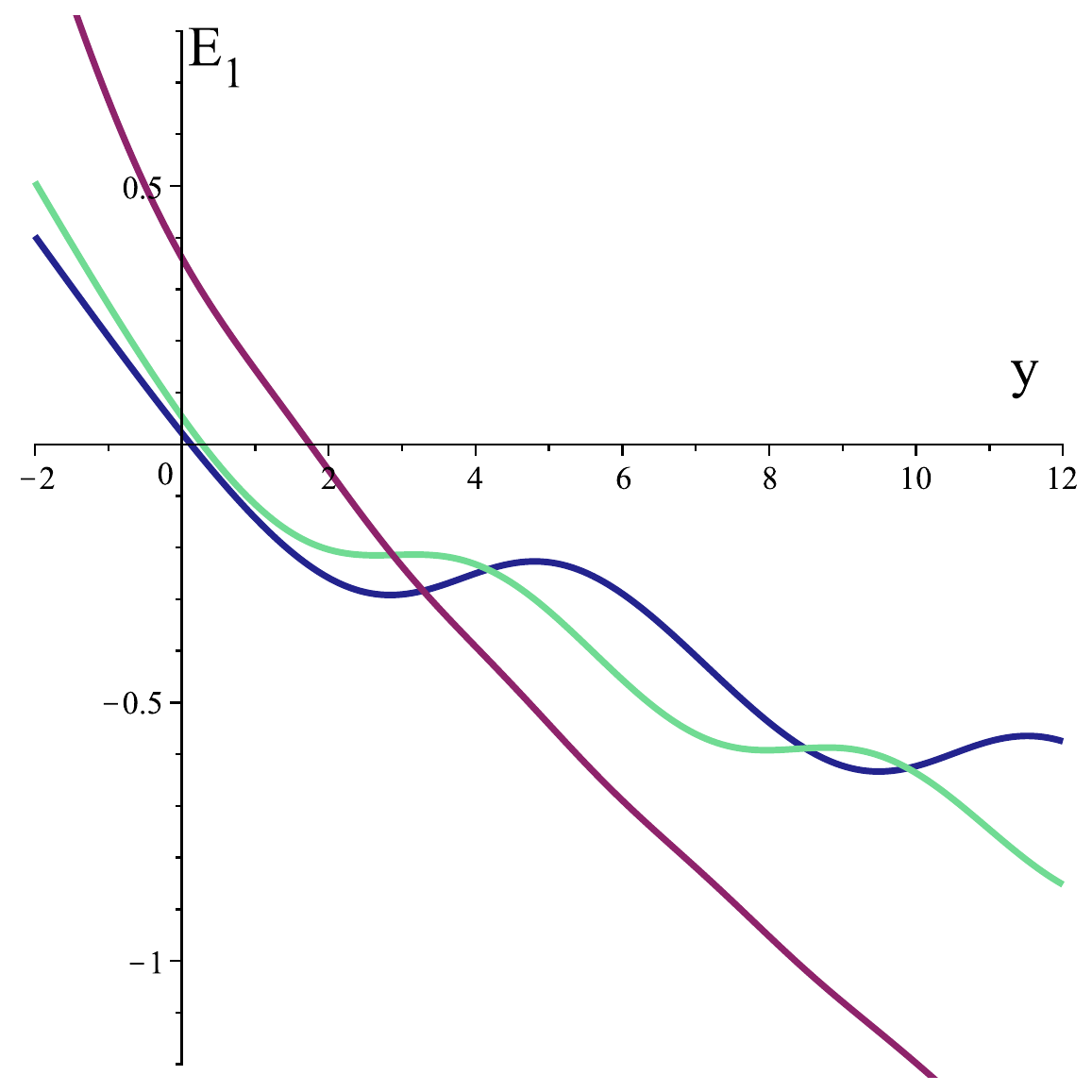}
	\hfill
	\includegraphics[width=0.3\textwidth,angle=0]{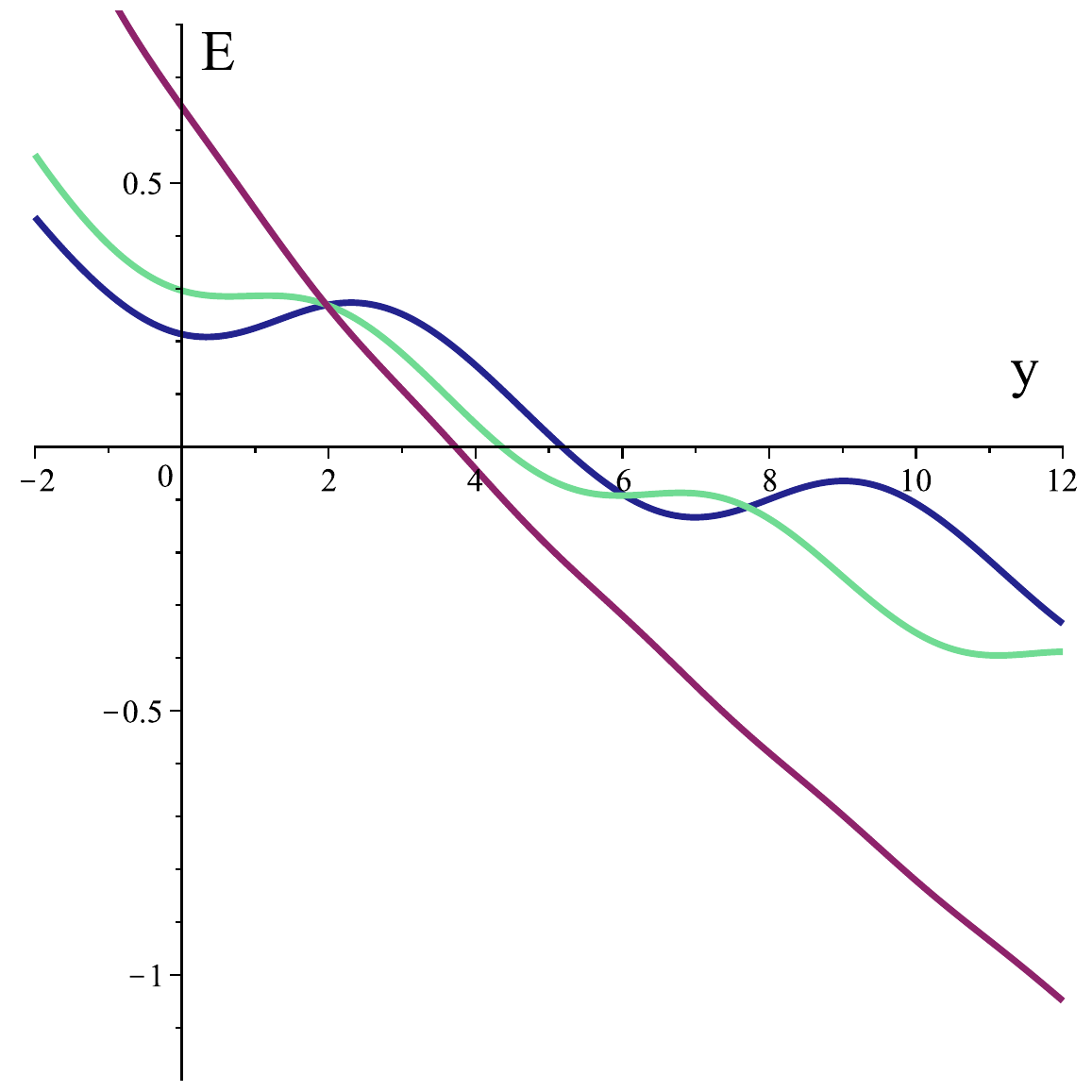}
	\vfill
	\parbox{0.95\textwidth}{\caption{\label{fig:E1y_vs_y_b} Quantity $E_1(T^*,\mu^*;y)$ as a function of $y$ at different values of $T^*$ and $\mu^*$. Left: $\mu^*=-1.0$; Center: $\mu^*=0.0$; Right: $\mu^*=0.5$. Curves: Navy: $T^*=0.2$; Aquamarine: $T^*=0.25$; Maroon: $T^*=0.5$.}}
	
\end{figure}

\pagebreak
\subsection{Figures for $E_2(T^*,\mu^*;y)$}
\label{sec:figE2}

\begin{figure}[htbp]
	\includegraphics[width=0.3\textwidth,angle=0]{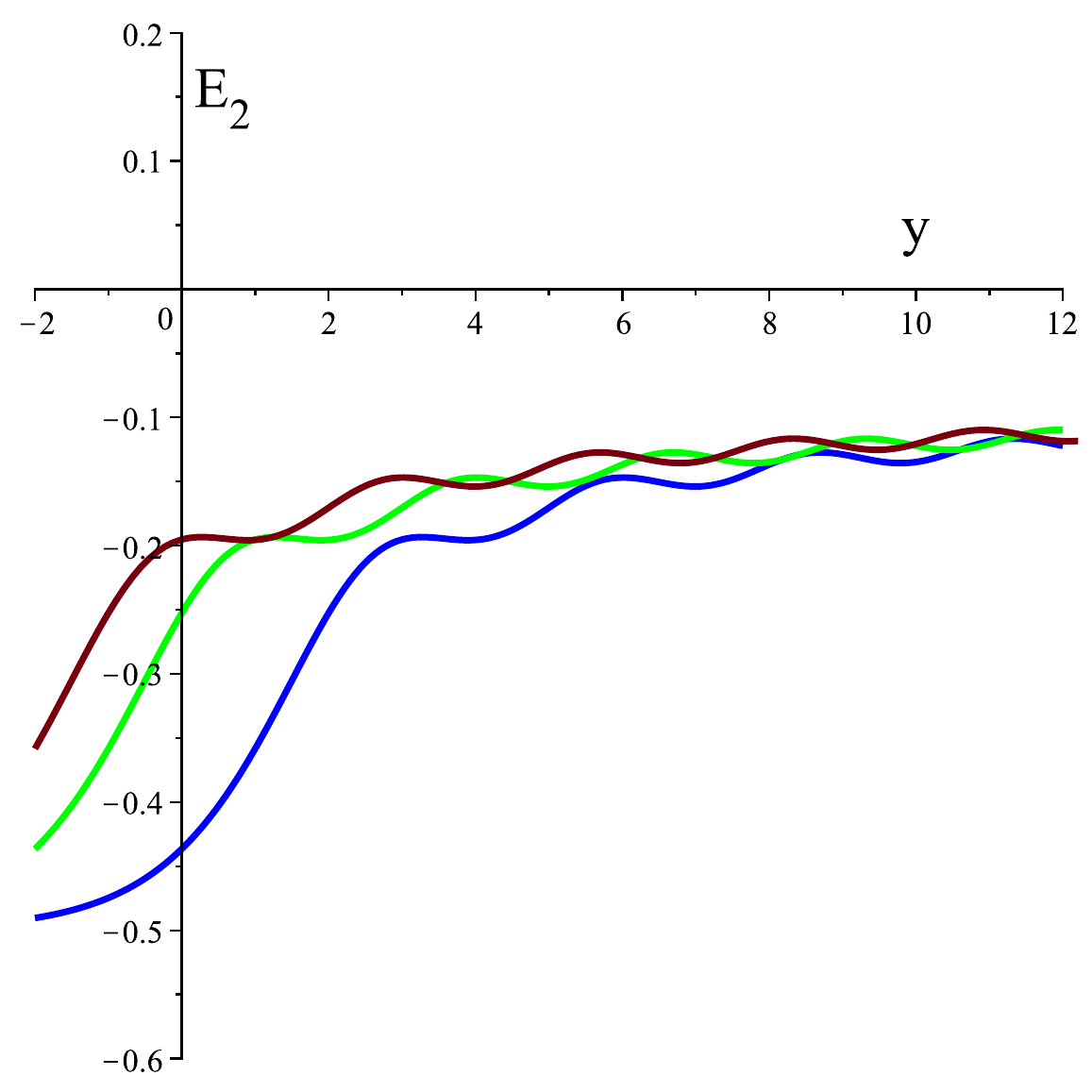}
	\hfill
	\includegraphics[width=0.3\textwidth,angle=0]{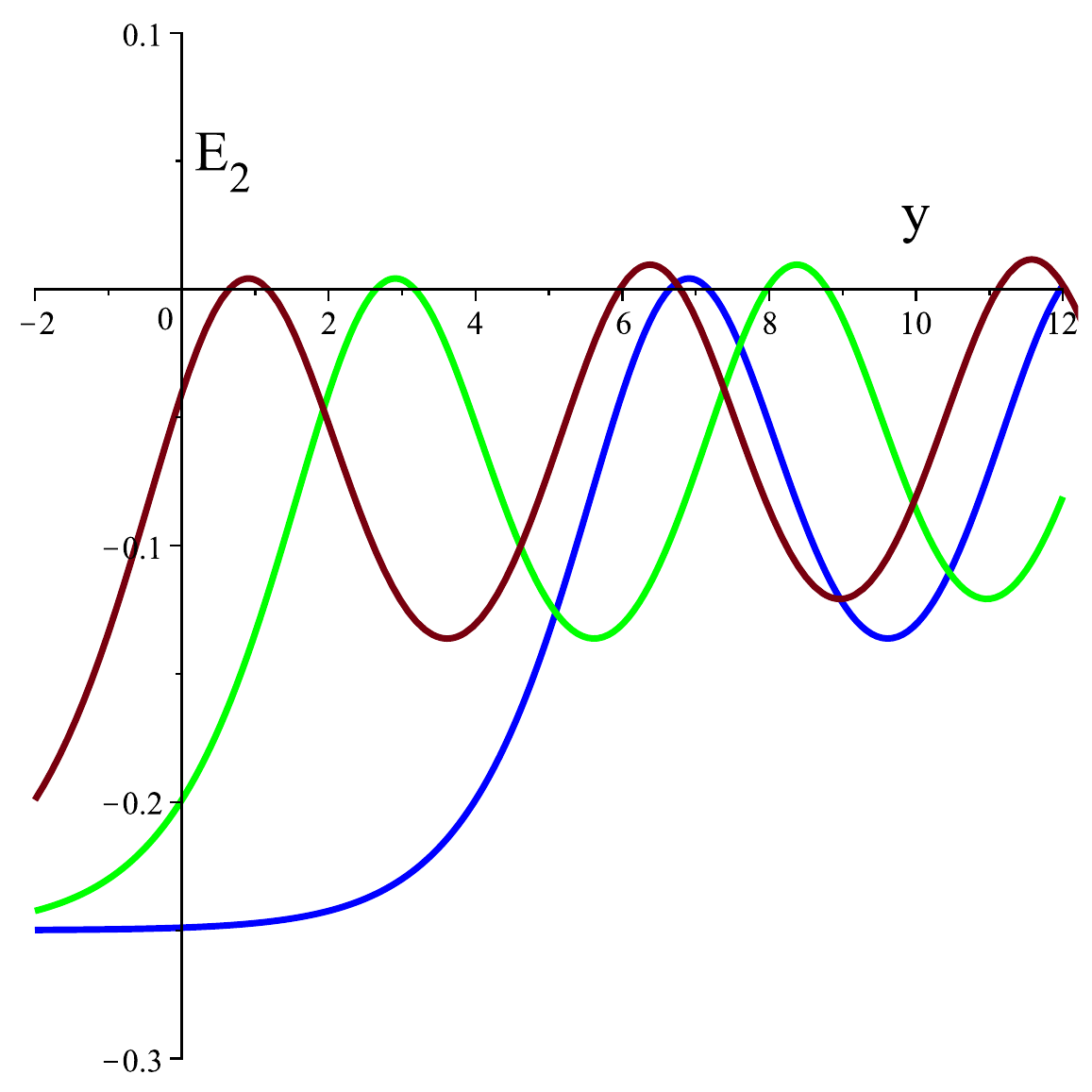}
	\hfill
	\includegraphics[width=0.3\textwidth,angle=0]{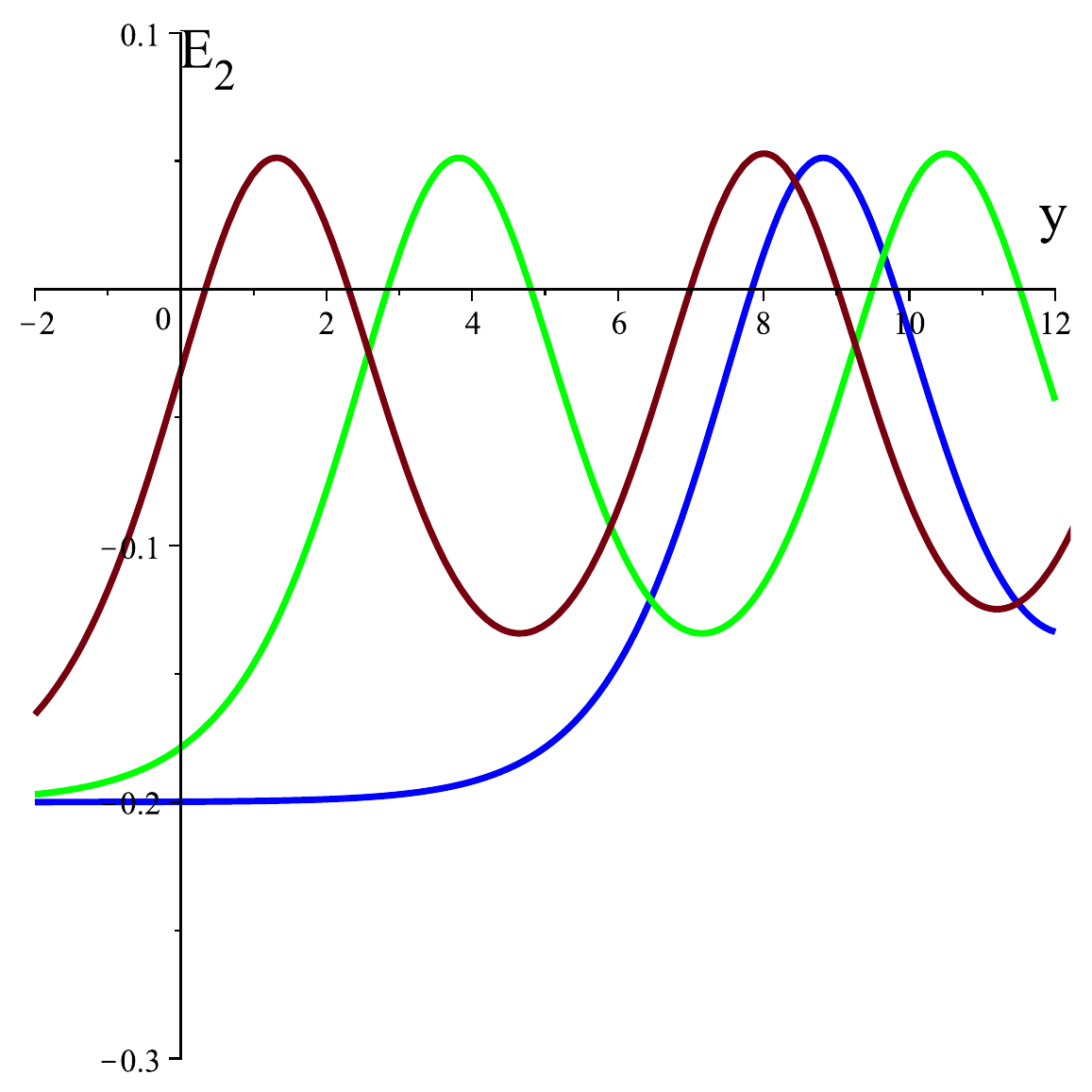}
	\\
	\vfill
	\parbox{0.95\textwidth}{\caption{\label{fig:E2y_vs_y_a} Quantity $E_2(T^*,\mu^*;y)$ as a function of $y$ at different values of $T^*$ and $\mu^*$. Left: $T^*=0.5$; Center: $T^*=0.25$; Right: $T^*=0.2$. Curves: Blue: $\mu^*=-1.0$; Green: $\mu^*=0.0$; Red: $\mu^*=0.5$.}}
	
\end{figure}

\begin{figure}[htbp]
	\includegraphics[width=0.3\textwidth,angle=0]{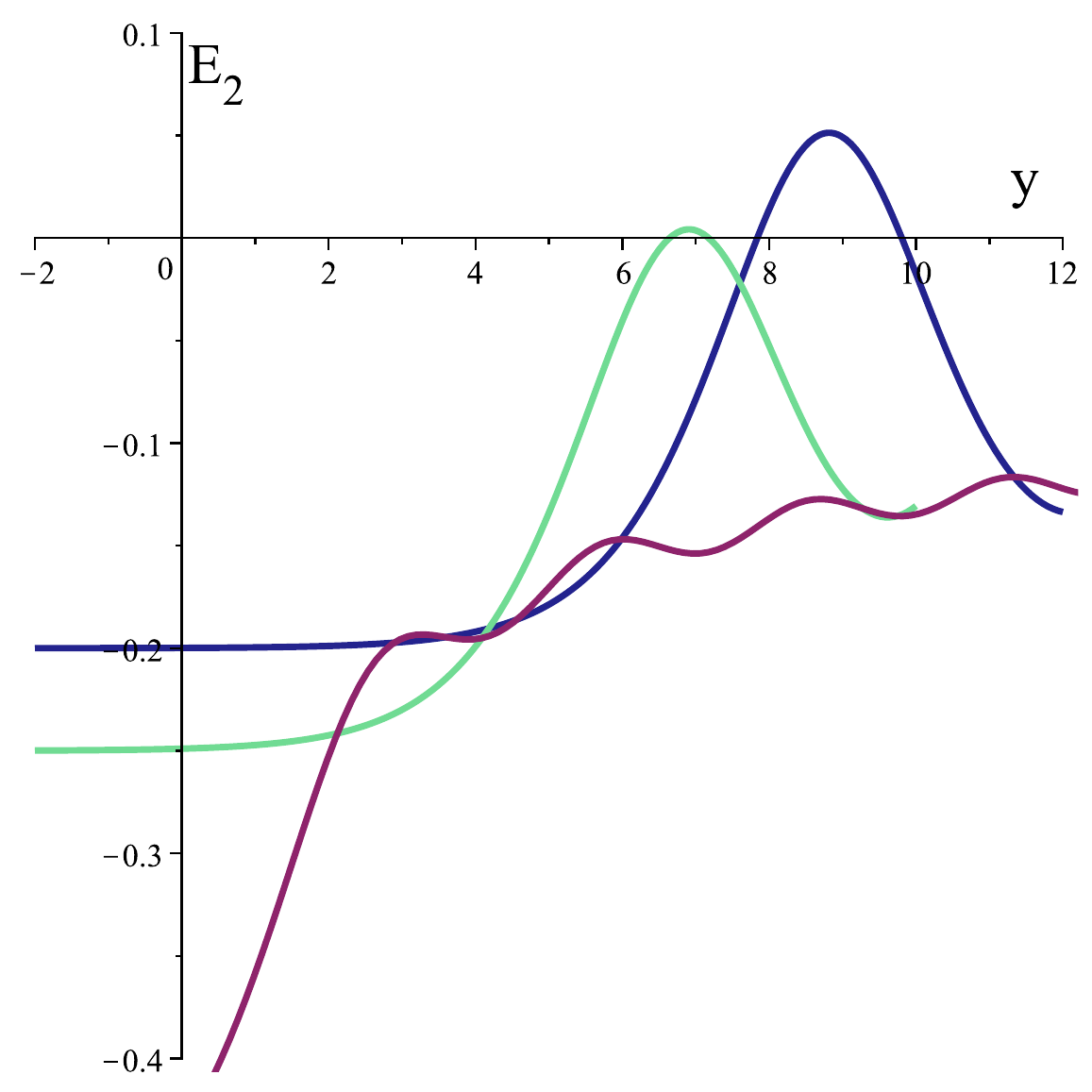}
	\hfill
	\includegraphics[width=0.3\textwidth,angle=0]{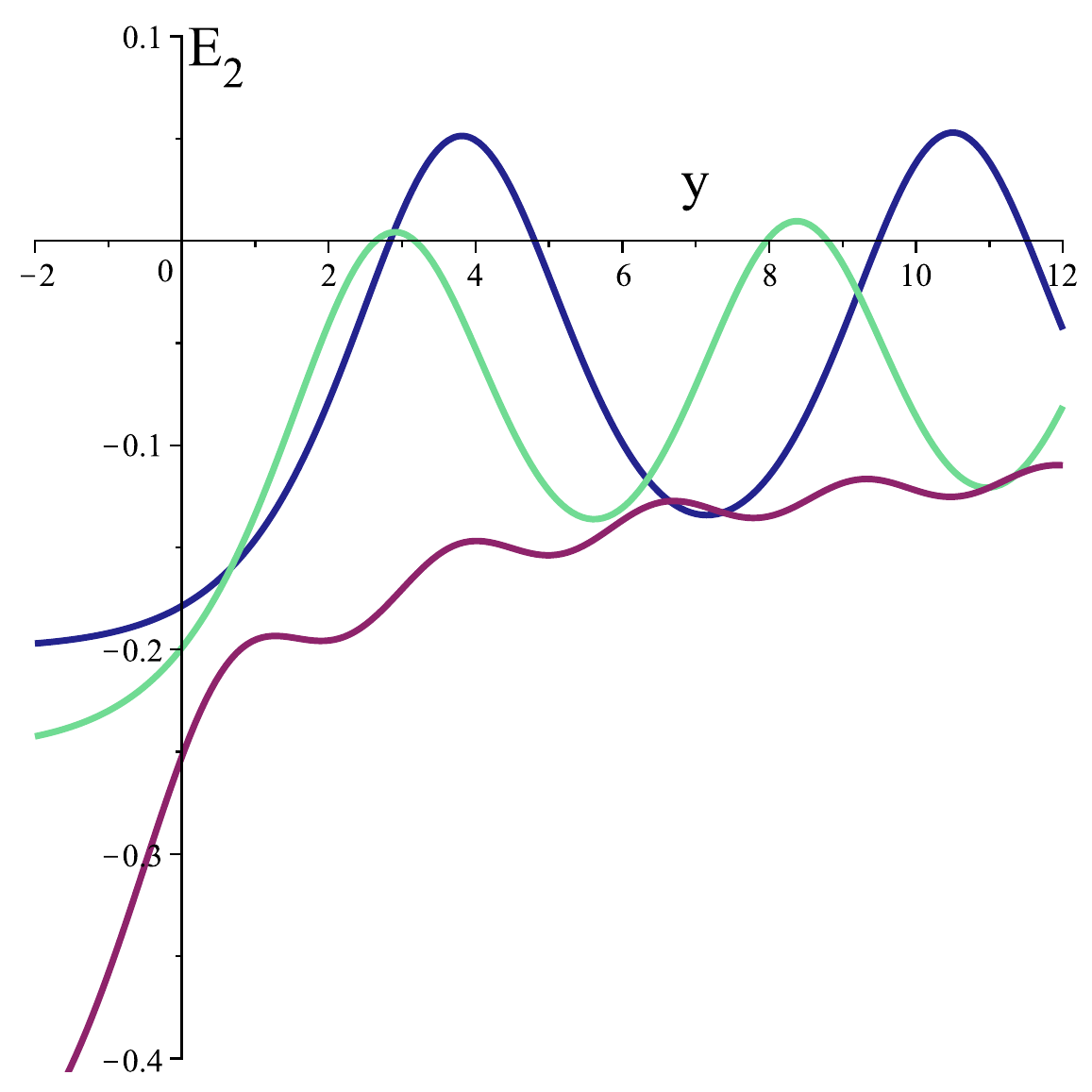}
	\hfill
	\includegraphics[width=0.3\textwidth,angle=0]{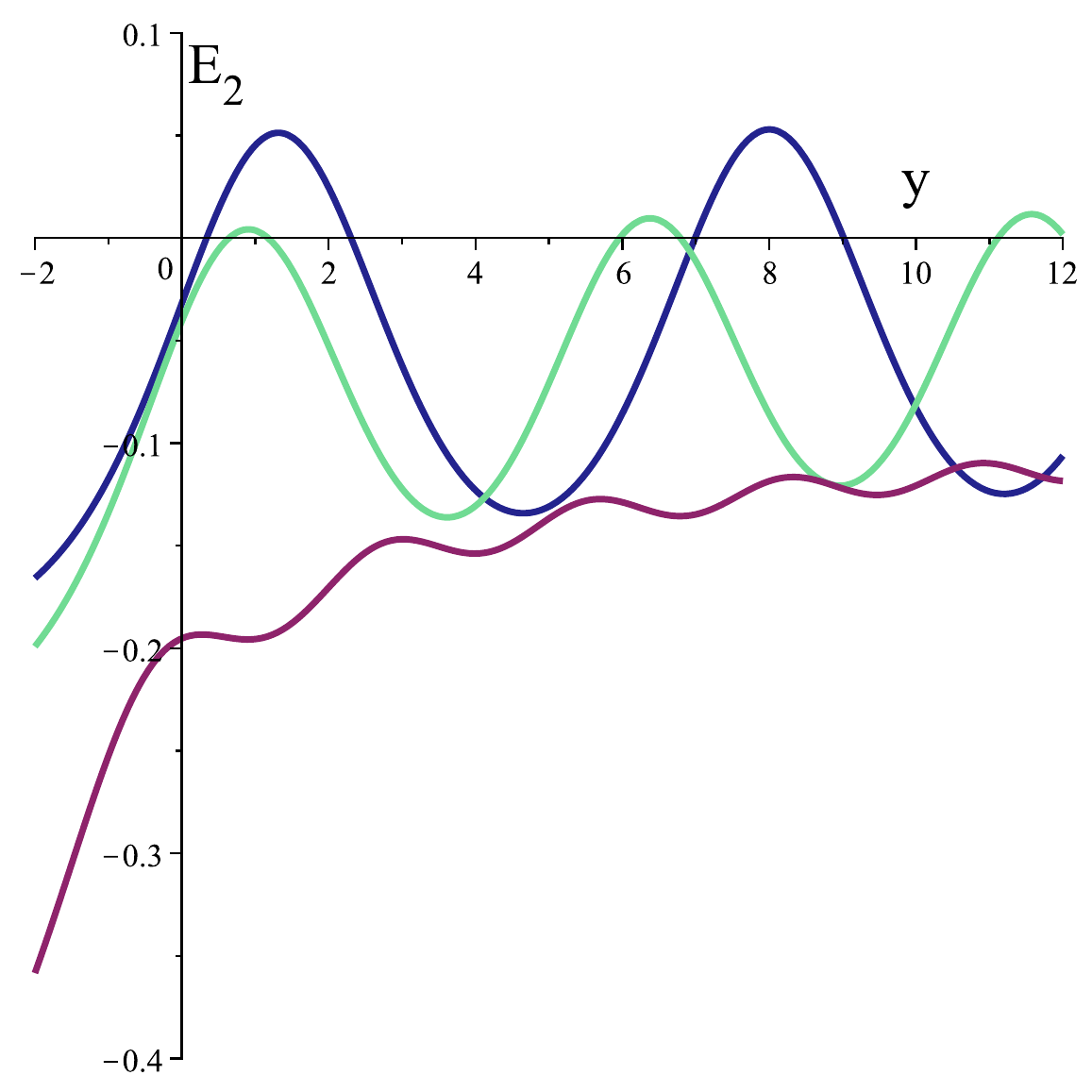}
	\vfill
	\parbox{0.95\textwidth}{\caption{\label{fig:E2y_vs_y_b} Quantity $E_2(T^*,\mu^*;y)$ as a function of $y$ at different values of $T^*$ and $\mu^*$. Left: $\mu^*=-1.0$; Center: $\mu^*=0.0$; Right: $\mu^*=0.5$. Curves: Navy: $T^*=0.2$; Aquamarine: $T^*=0.25$; Maroon: $T^*=0.5$.}}
	
\end{figure}

\pagebreak

\section{Critical-point conditions}\label{CPC}

In this appendix, we calculate the first and second derivatives of the pressure with respect to the density, and establish the connection between the general critical-point conditions~\eqref{split} and the equations~\eqref{eq:system}
specific for the system considered throughout the paper.

For the first derivative, we obtain
\begin{eqnarray}
	\label{sp1_b}
	\left(\frac{\partial P^*}{\partial \rho^*}\right)_{T^*} & = &
	\frac{\left(\partial P^* / \partial \mu^*\right)_{T^*}}{\left(\partial \rho^* / \partial \mu^* \right)_{T^*}}
	= \frac{\rho^*(T^*,\mu^*)}{T^* \left(\partial \bar{y}_{\rm max} / \partial \mu^*\right)_{T^*}},
\end{eqnarray}
where we applied \eqref{eq:dens} and \eqref{rho_vs_T_mu}.

The first condition from \eqref{cond:max}
\begin{equation}\label{sp3_b}
	E_1\left(T^*,\mu^*;\bar{y}_{\rm max}\right)=0\,,
\end{equation}
implicitly defines a functional relation between $\mu^*$ and $\bar{y}_{\rm max}$ at fixed temperature $T^*$, as required in \eqref{split} and \eqref{sp1_b}. Implicit differentiation of \eqref{sp3_b} allows us to determine the derivative $\partial \bar{y}_{\rm max} / \partial \mu^*$ in \eqref{sp1_b} via 
(see e.g.~\cite[p.~8]{KrantzParks03}, \cite[Appendix A-5]{Callen85})
\begin{equation}
	\left(\frac{\partial \bar{y}_{\rm max}}{\partial \mu^*}\right)_{T^*} =
	- \left(\frac{\partial E_1(T^*,\mu^*; \bar{y}_{\rm max})}{\partial \mu^*} \right)_{T^*,\bar{y}_{\rm max}} 
	\left[ \left(\frac{\partial E_1(T^*,\mu^*; \bar{y}_{\rm max})}{\partial \bar{y}_{\rm max}}\right)_{T^*,\mu^*}\right]^{-1}.
\end{equation}
Direct differentiation in the second factor immediately yields $\partial E_1 / \partial \bar{y}_{\rm max} = E_2(T^*,\mu^*; \bar{y}_{\rm max})$. For the first factor, the result is also straightforward, taking into account \eqref{def:reducedE1} and \eqref{def:reducedE2}:
\begin{equation}
	\left(\frac{\partial E_1(T^*,\mu^*; \bar{y}_{\rm max})}{\partial \mu^*}\right)_{T^*, \bar{y}_{\rm max}} = \frac{1}{T^*} \left[T^* + E_2(T^*,\mu^*; \bar{y}_{\rm max})\right].
\end{equation}
Substituting these results back into \eqref{sp1_b}, we arrive at
\begin{equation}\label{sp7_b}
	\left(\frac{\partial P^*}{\partial \rho^*}\right)_{T^*}=
	-\rho^*\,\frac{E_2(T^*,\mu^*;\bar{y}_{\rm max})}
	{T^*+E_2\left(T^*,\mu^*;\bar{y}_{\rm max}\right)}=0\,.
\end{equation}
Given that $\rho^*$ and $T^*+E_2\left(T^*,\mu^*;\bar{y}_{\rm max}\right)$ are finite (see \eqref{dE2}), the last condition implies that $E_2(T^*,\mu^*;\bar{y}_{\rm max})=0$ at the critical point:
\begin{equation}
	\label{C2}
	\left(\frac{\partial P^*}{\partial \rho^*}\right)_{T^*} = 0 \implies E_2 = 0.
\end{equation}

For further calculations, the following result is useful
\begin{equation}
	\left( \frac{\partial \rho^*}{\partial \mu^*} \right)_{T^*} = T^* \left( \frac{\partial \bar{y}_{\rm max}}{\partial \mu^*} \right)_{T^*}
	= -\frac{E_2 + T^*}{E_2}.
\end{equation}
Now, the second derivative of the pressure with respect to density is calculated as follows
\begin{eqnarray}
	\left(\frac{\partial^2 P^*}{\partial \rho^{*2}}\right)_{T^*} & = & -\frac{\partial}{\partial \rho^*} \left(\frac{K_1}{K_0} \frac{E_2}{E_2 + T^*} \right)_{T^*}
	\nonumber\\
	& = & -\frac{\partial}{\partial \mu^*} \left(\frac{K_1}{K_0} \frac{E_2}{E_2 + T^*} \right)_{T^*} \bigg/ \left(\frac{\partial \rho^*}{\partial \mu^*}\right)_{T^*}
	\nonumber\\
	& = & \frac{E_2}{E_2 + T^*} \left[- \frac{K_2K_0 - K_1^2}{K_0^2} \frac{1}{E_2 + T^*} + \frac{K_1}{K_0} \frac{E_3}{(E_2 + T^*)^2}\right]
	\nonumber\\
	& & - \frac{K_1}{K_0} \frac{E_3}{(E_2 + T^*)^2},
\end{eqnarray}
where we implied that each quantity $K_j$ and $E_j$ is a function of arguments $(T^*,\mu^*;\bar{y}_{\rm max})$.

Therefore, taking into account Eq.~\eqref{C2}, we conclude
\begin{equation}
	\left(\frac{\partial^2 P^*}{\partial \rho^{*2}}\right)_{T^*} = 0 \implies E_3 = 0,
\end{equation}
which is the last entry in the system of equations~\eqref{eq:system}.

\section{\label{sec:app:entropy} Calculation of Entropy}
This Appendix contains explicit calculation of entropy from Eq.~\eqref{eq:entropy} by taking derivatives of pressure with respect to temperature
\begin{equation}
	S^{*} = \frac{1}{\rho^*} \left[E + T^* \left(\frac{\partial E}{\partial T^*}\right)_{\mu^*}\right];
\end{equation}
\begin{equation}
	\frac{\partial E}{\partial T^*} = -\frac{1}{2}\bar{y}_{\rm max}^2 - T^* \bar{y}_{\rm max} \frac{\partial \bar{y}_{\rm max}}{\partial T^*} + \frac{1}{K_0}\frac{\partial}{\partial T^*} K_0.
\end{equation}
\begin{eqnarray}
	\frac{\partial K_0}{\partial T^*} & = & \sum_{n=0}^{\infty} \frac{n v^*(v^* T^{*3/2})^{n-1}}{n!} \frac{3T^{*1/2}}{2}\exp(F_n)
	\nonumber\\
	& + & \sum_{n=0}^{\infty} \frac{(v^* T^{*3/2})^n}{n!} \exp(F_n) \frac{\partial}{\partial T^*} F_n,
\end{eqnarray}
where we temporary introduced
\begin{eqnarray*}
	F_n & = & \left(\bar{y}_{\rm max} + \frac{\mu^*}{T^*} \right)n -\frac{a}{2T^*}n^2.
\end{eqnarray*}
The derivative of $F_n$ with respect to $T^*$ is
\begin{equation}
	\frac{\partial F_n}{\partial T^*} = -\frac{1}{T^*} \left(\frac{\mu^*}{T^*}n - \frac{a}{2T^*}n^2\right) + n \frac{\partial \bar{y}_{\rm max}}{\partial T^*}.
\end{equation}
Substituting the last result into the derivative for $K_0$, we obtain
\begin{equation}
	\frac{\partial K_0}{\partial T^*} = \frac{1}{T^*} \left(\frac{3}{2} - \frac{\mu^*}{T^*} \right)K_1 + \frac{1}{T^*}\frac{a}{2 T^*} K_2 + K_1 \frac{\partial \bar{y}_{\rm max}}{\partial T^*}.
\end{equation}
For the derivative of $E$ with respect to $T^*$ we have
\begin{eqnarray}
	\frac{\partial E}{\partial T^*} & = & -\frac{1}{2}\bar{y}_{\rm max}^2 + \frac{1}{T^* K_0} \left[\left(\frac{3}{2} - \frac{\mu^*}{T^*} \right)K_1 + \frac{a}{2 T^*} K_2\right] + \left(-T^*\bar{y}_{\rm max} + \frac{K_1}{K_0}\right) \frac{\partial \bar{y}_{\rm max}}{\partial T^*}
	\nonumber\\
	& = & -\frac{1}{2}\bar{y}_{\rm max}^2 + \frac{1}{T^* K_0} \left[\left(\frac{3}{2} - \frac{\mu^*}{T^*} \right)K_1 + \frac{a}{2 T^*} K_2\right],
\end{eqnarray}
where we accounted for~\eqref{eq:bar_y}.

The final result for the entropy per particle is
\begin{eqnarray}
	S^{*} 
	& = & \frac{1}{\rho^*} \left\{ -\frac{1}{T^*} \frac{K_1^2}{K_0^2} + \ln K_0 + \frac{1}{K_0} \left[\left(\frac{3}{2} - \frac{\mu^*}{T^*} \right)K_1 + \frac{a}{2 T^*} K_2\right] \right\}.
\end{eqnarray}
Taking into account~\eqref{eq:densK}, we rewrite the final result as
\begin{equation}
	S^* = \left(\frac{3}{2} - \frac{\mu^*}{T^*}\right) - \frac{1}{T^*}\frac{K_1}{K_0} + \frac{K_0 \ln K_0}{K_1} + \frac{a}{2T^*} \frac{K_2}{K_1}.
\end{equation}

Note that we may have ignored here the dependence of $\bar{y}_{\rm max}$ on temperature, since $\bar{y}_{\rm max}$ maximizes $E$ and thus the contribution of partial derivative of $E$ with respect to $\bar{y}_{\rm max}$ will be zero. However, such dependence must be accounted when calculating higher order derivatives, e.g. when calculating the heat capacity $C_V$.




\end{appendices}


\bibliography{references}
\addcontentsline{toc}{section}{References}

\end{document}